\documentclass[letterpaper,12pt]{article}

\pdfoutput=1
\usepackage{graphicx}
\usepackage{amsmath}
\usepackage{hyperref}
\usepackage{bm}

\makeatletter
\renewcommand\section{\@startsection {section}{1}{\z@}%
{-3.5ex \@plus -1ex \@minus -0.2ex}%
{2.3ex \@plus 0.2ex}%
{\normalfont\normalsize\bfseries}}

\renewcommand\subsection{\@startsection{subsection}{2}{\z@}%
{-3.25ex \@plus -1ex \@minus -0.2ex}%
{1.5ex \@plus 0.2ex}%
{\normalfont\normalsize\bfseries}}

\def\@seccntformat#1{\csname the#1\endcsname.\quad}
\makeatother

\newcommand{\oline}[1]{\overline{\mkern-1.0mu#1\mkern0.0mu}}
\newcommand{\overbar}[1]{\mkern 4.3mu\overline{\mkern-4.3mu#1\mkern-1.5mu}\mkern 1.5mu}
\newcommand\Fprime{F\hspace*{0.05em}^{\prime}}

\begin{document}

\setlength{\baselineskip}{4.5ex}

\noindent
{\bf \LARGE Sharp hypotheses and bispatial inference}\\[3ex]

\noindent
{\bf Russell J. Bowater}\\
\emph{Independent researcher, Sartre 47, Acatlima, Huajuapan de Le\'{o}n, Oaxaca, C.P.\ 69004,
Mexico. Email address: as given on arXiv.org. Twitter profile:
\href{https://twitter.com/naked_statist}{@naked\_statist}\\ Personal website:
\href{https://sites.google.com/site/bowaterfospage}{sites.google.com/site/bowaterfospage}}
\\[2ex]

\noindent
{\small \bf Abstract:}
{\small
A fundamental class of inferential problems are those characterised by there having been a
substantial degree of pre-data (or prior) belief that the value of a model parameter was equal or
lay close to a specified value, which may, for example, be the value that indicates the absence of
an effect.
Standard ways of tackling problems of this type, including the Bayesian method, are often highly
inadequate in practice.
To address this issue, an inferential framework called bispatial inference is put forward, which
can be viewed as both a generalisation and radical reinterpretation of existing approaches to
inference that are based on P values. 
It is shown that to obtain an appropriate post-data density function for a given parameter, it is
often convenient to combine a special type of bispatial inference, which is constructed around
one-sided P values, with a previously outlined form of fiducial inference.
Finally, by using what are called post-data opinion curves, this bispatial-fiducial theory is
naturally extended to deal with the general scenario in which any number of parameters may be
unknown.
The application of the theory is illustrated in various examples, which are especially relevant to
the analysis of clinical trial data.}
\\[3ex]

\pagebreak
\section{Introduction}

Let us imagine that our aim is to make inferences about an unknown parameter $\nu$ on the basis of
a data set $x$ that was generated by a sampling model that depends on the true value of $\nu$.
Given this context, we will begin with the following definition.

\vspace*{2ex}
\noindent
{\bf Definition 1: Sharp and almost sharp hypotheses}

\vspace*{1ex}
\noindent
The hypothesis that the parameter $\nu$ lies in an interval $[\nu_0, \nu_1]$ will be defined as a
sharp hypothesis if $\nu_0 = \nu_1$, and as an almost sharp hypothesis if the difference $\nu_1 -
\nu_0$ is very small in the context of our general uncertainty about $\nu$ after the data $x$ have
\linebreak been observed.

\vspace*{2ex}
Clearly, any importance attached to a hypothesis of either of these two types should not generally
have a great effect on the way that we make inferences about $\nu$ on the \linebreak basis of the
data $x$ if there had been no exceptional reason to believe that it would have been true or false
before the data were observed. Taking this into account, it will be assumed that we are in the
following scenario.

\vspace*{2ex}
\noindent
{\bf Definition 2: Scenario of interest}

\vspace*{1ex}
\noindent
This scenario is characterised by there having been a substantial degree of belief before the data
were observed, i.e.\ a substantial pre-data belief, that a given sharp or almost sharp hypothesis
about the parameter $\nu$ could have been true, but if, on the other hand, this hypothesis had
been conditioned to be false, i.e.\ if $\nu$ had been conditioned not to lie in the interval
$[\nu_0, \nu_1]$, then there would have been very little or no pre-data knowledge about
this parameter over all of its allowable values outside of this interval.
In this scenario, the hypothesis in question will be referred to as the \emph{special} hypothesis.

\vspace*{2ex}
Perhaps some may try to dismiss the importance of this type of scenario, however trying to make
data-based inferences about any given parameter of interest $\nu$ in such a scenario represents one
of the most fundamental problems of statistical inference that arise in practice. Let us consider
the following examples.

\vspace*{2ex}
\noindent
{\bf Example 1: Intervening in a system}

\vspace*{1ex}
\noindent
If $\nu$ is a parameter of one part of a system, and an intervention is made in a second part of
the system that is arguably completely disconnected from the first part, then there will be a high
degree of belief that the value of $\nu$ will not change as a result of the intervention, i.e.\
there is a strong belief in a sharp hypothesis about $\nu$.

\vspace*{2ex}
\noindent
{\bf Example 2: A randomised-controlled trial}

\vspace*{1ex}
\noindent
Let us imagine that a sample of patients is randomly divided into a group of $n_t$ patients, namely
the treatment group, that receive a new drug B, and a group of $n_c$ patients, namely the control
group, that receive a standard drug A.
We will assume that $e_t$ patients in the treatment group experience a given adverse event, e.g.\ a
heart attack, in a certain period of time following the start of treatment, and that $e_c$ patients
in the control group experience the same type of event in the same time period.
On the basis of this sample information, it will be supposed that the aim is to make inferences
about the relative risk $\pi_t/\pi_c$, where $\pi_t$ and $\pi_c$ are the population proportions of
patients who would experience the adverse event when given drug B and drug A respectively.
Now, if the action of drug B on the body is very similar to the action of drug A, which is in fact
often the case in practice when two drugs are being compared in this type of clinical trial, then
there may well have been a strong pre-data belief that this relative risk would be close to one, or
in other words, that the almost sharp hypothesis that the relative risk would lie in a narrow
interval containing the value one would be true.

\vspace*{2ex}
It would appear that a common way to deal with there having been a strong pre-data belief that a
sharp or almost sharp hypothesis was true is to simply ignore the inconvenient presence of this
belief.
However, doing so means that inferences based on the observed data will often not be even remotely
honest. On the other hand, a formal method of addressing this issue that has received some
attention is the Bayesian method. Let us take a quick look at how this method would work in a
simple example.

\vspace*{2ex}
\noindent
{\bf Example 3: Application of the Bayesian method}

\vspace*{1ex}
\noindent
Let us suppose that we are interested in making inferences about the mean $\mu$ of a normal density
function that has a known variance $\sigma^2$, on the basis of a sample of values $x$ drawn from
the density function concerned. It will be assumed that we are in the scenario of Definition~2 with
the special hypothesis of interest being the sharp hypothesis that $\mu=0$. Under the Bayesian
paradigm, it would be natural to incorporate any degree of pre-data belief that $\mu$ equals zero
into the analysis of the data by assigning a positive prior probability to this hypothesis.

However, the only accepted way of expressing a lack of knowledge about a model parameter under this
paradigm is the controversial strategy of placing a diffuse proper or improper prior density over
the parameter concerned. Taking this into account, let us assume, without a great loss of
generality, that the prior density function of $\mu$ conditional on $\mu \neq 0$ is a normal
density function with a mean of zero and a large\linebreak variance $\sigma_{0}^2$.

The inadequacy of the strategy in question is clearly apparent in the uncertainty there would be in
choosing a value for the variance $\sigma_{0}^2$, and this issue becomes very hard to conceal after
appreciating that the amount of posterior probability given to\linebreak the hypothesis that
$\mu=0$ is highly sensitive to changes in this variance. For example, the natural desire to allow
the variance $\sigma_{0}^2$ to tend to infinity results in the posterior probability of this
hypothesis tending to one for any given data set $x$ and any given positive prior probability that
is assigned to $\mu$ equalling zero.

\vspace*{2ex}
It can be easily argued, therefore, that the application of standard Bayesian theory in the case
just examined has an appalling outcome.
Moreover, applying the Bayesian strategy just described leads to outcomes of a similar type in
cases where the sampling density of the data given the parameter of interest $\nu$ is not normal,
and/or the prior density of this parameter has a more general form, and also, importantly, in cases
where the special hypothesis is an almost sharp rather than, simply, a sharp hypothesis.
This clearly gives us a strong motivation to look for an alternative method for making inferences
about $\nu$ in the scenario of interest.
Following a similar path to that of Bowater and Guzm\'an-Pantoja~(2019b), the aim of the present
paper is to develop a satisfactory method for doing this on the basis of classical ideas about
statistical inference. This method of inference will be called bispatial inference.

Before going further, let us summarise the structure of the paper. In the next section, a general
theory of bispatial inference is broadly outlined. A special formalisation of this theory is then
developed in detail in Section~\ref{sec19}. Given that all reasonable objectives for making
inferences about a parameter of interest $\nu$ can not be conveniently achieved by using this
theory alone, a method of inference is put forward in Section~\ref{sec11} that is based on
combining bispatial inference with a specific type of fiducial inference. In the final main section
of the paper, namely Section~\ref{sec20}, this combined theory is extended to cases where various
model parameters are unknown.

\vspace*{3ex}
\section{General theory of bispatial inference}

\vspace*{1ex}
\subsection{Overall problem}
\label{sec24}

Let us now consider a more general problem of statistical inference to the one that was discussed
in the Introduction. In particular, we now will be interested in the problem of making inferences
about a set of parameters $\theta =\{\theta_i : i=1,2,\ldots,k\}$, where each $\theta_i$ is a
one-dimensional variable, on the basis of a data set $x=\{x_i : i=1,2,\ldots,n\}$ that was
generated by a sampling model that depends on this set of parameters. Let the joint density or mass
function of the data given the true values of the parameters $\theta$ be denoted as
$g(x\,|\,\theta)$. This will be the overall problem of inference that we will be concerned with in
the rest of this paper.

\vspace*{3ex}
\subsection{A note about probability}
\label{sec3}

We will interpret the concept of probability under the definition of generalised subjective
probability that was comprehensively outlined in Bowater~(2018b).
Given that it will not be necessary to explicitly discuss this definition of probability in the
present paper, the reader is referred to this earlier work for further information.
Nevertheless, in relation to the general topic in question, there is a specific issue that should
not be overlooked.
In particular, we observe that when events are repeatable, the concept of the probability of an
event and the concept of the proportion of times the event occurs in the long term are often used
interchangeably.
However, this is not always appropriate.
The reason for this is that a population proportion is a fact about the physical world, while under
certain definitions of probability, e.g.\ the definition that will be adopted here, a probability
is primarily always a measure of a given individual's state of mind.
Therefore, where necessary, we will denote the population proportion of times any given event $A$
occurs by $\rho(A)$, while the probability of the event $A$ will, as usual, be denoted by $P(A)$.

\vspace*{3ex}
\subsection{Parameter and sampling space hypotheses}
\label{sec2}

The theory of inference that will be developed is based on a hypothesis $H_{P}$ that concerns an
event in the parameter space, and an equivalent hypothesis $H_{S}$ that is stated in terms of the
proportion of times an event in the sampling space will occur in the long run. The link that is
made between the parameter and sampling spaces through the attention given to these two hypotheses
is the reason that this type of inference will be called \emph{bispatial} inference.
More specifically, these two types of hypothesis will be assumed to have the following definitions.

\vspace*{2ex}
\noindent
{\bf Definition~3: Parameter space hypothesis $H_{P}$}

\vspace*{1ex}
\noindent
Given that, from now, $H\,$:$\,C$ will denote the hypothesis $H$ that a given condition $C$ is
true, the parameter space hypothesis $H_{P}$ is defined by:
\vspace*{-0.75ex}
\[
H_{P}: \theta \in \Theta_0
\vspace*{-0.75ex}
\]
where $\Theta_0$ is a given subset of the entire space $\Theta$ over which the set of parameters
$\theta$ is defined.

\vspace*{2ex}
\noindent
{\bf Definition~4: Sampling space hypothesis $H_{S}$}

\vspace*{1ex}
\noindent
The two conditions that the sampling space hypothesis $H_{S}$ must satisfy are:

\vspace*{1ex}
\noindent
1) It must be equivalent to the hypothesis $H_{P}$, i.e.\ if $H_{S}$ is true then $H_{P}$ must be
true and if $H_{P}$ is true then $H_{S}$ must be true.

\vspace*{1ex}
\noindent
2) It must have the following form:
\vspace*{-0.25ex}
\[
H_{S}: \rho (J(X^*) \in \mathcal{J}_0(x)) \in \mathcal{P}_0
\vspace*{-0.25ex}
\]
where $J(X^*)$ is a statistic calculated on the basis of an as-yet-unobserved second sample $X^*$
of values drawn from the density function $g(x\,|\,\theta)$, which is possibly of a different size
to the observed (first) sample $x$, the set $\mathcal{J}_0(x)$ is a given subset of the entire
space $\mathcal{J}$ over which the statistic $J(X^*)$ is defined, and the set $\mathcal{P}_0$ is a
given subset of the interval $[0,1]$. To clarify, the hypothesis $H_{S}$ is the hypothesis that the
unknown population proportion $\rho (J(X^*) \in \mathcal{J}_0(x))$ lies in the known set
$\mathcal{P}_0$. Also, it should be clarified that the definition of the set $\mathcal{J}_0(x)$
will depend, in general, on the data set $x$.

\pagebreak
\subsection{Inferential process}
\label{sec1}

It will be assumed that inferences are made about the set of parameters $\theta$ by proceeding
through the steps of the following algorithm:

\vspace*{1.5ex}
\noindent
Step 1: Formation of a suitable hypothesis $H_{P}$. The choice of this hypothesis should be made
with the goal in mind of being able to make useful inferences about the parameters~$\theta$.

\vspace*{1.5ex}
\noindent
Step 2: Assessment of the likeliness of the hypothesis $H_{P}$ being true using only pre-data
knowledge about the parameters $\theta$. It is not necessary that this assessment is expressed in
terms of a formal measure of uncertainty, e.g.\ a probability does not need to be assigned to this
hypothesis.

\vspace*{1.5ex}
\noindent
Step 3: Formation of a suitable hypothesis $H_{S}$.

\vspace*{1.5ex}
\noindent
Step 4: Assessment of the likeliness of the hypothesis $H_{S}$ being true after the data $x$ have
been observed. In carrying out this assessment, all relevant factors ought to be taken into account
including, in particular, the assessment made in Step 2 and the known equivalency between the
hypotheses $H_{P}$ and $H_{S}$.

\vspace*{1.5ex}
\noindent
Step 5: Conclusion about the likeliness of the hypothesis $H_{P}$ being true having taken into
account the data $x$. This is directly implied by the assessment made in Step~4 due to the
equivalence of the hypotheses $H_{P}$ and $H_{S}$.

\vspace*{3ex}
\subsection{First example: Two-sided P values}
\label{sec12}

In the next three sections, we will apply the method outlined in the previous section to the
problem of inference referred to in Example 3 of the Introduction, i.e.\ that of making inferences
about a normal mean $\mu$ when the population variance $\sigma^2$ is known. In this case, it is
clear that the set of unknown parameters $\theta$ will consist of just the mean $\mu$.

To give a context to this problem, let us imagine that a patient is being constantly monitored with
regard to the concentration of a certain chemical in his/her blood. We will assume that the
measurements of this concentration are notably imprecise, and in particular, it will be assumed
that any such measurement follows a normal density function with known variance $\sigma^2/2$
centred at the true concentration. Also, let us sup\-pose that the data $x$ is simply the
measurement of this concentration at a time point $\mathtt{t}_2$ minus the same type of measurement
taken at a time point $\mathtt{t}_1$, where the time point $\mathtt{t}_1$ is immediately before the
patient is subjected to some kind of intervention and the time point $\mathtt{t}_2$ is immediately
after this intervention.

Now, if the intervention in question would not be expected to affect the concentration of the
chemical of interest, there is likely to be a substantial degree of pre-data belief that the true
change in this concentration in going from the time point $\mathtt{t}_1$\hspace*{-0.05em} to the
time point $\mathtt{t}_2$, namely the change $\mu$ in this concentration, will be very small.
In fact, to begin with, let us assume that these two time points are so close together that we find
ourselves in the scenario of Definition~2 with the special hypothesis being the sharp hypothesis
that $\mu=0$. It can be seen therefore that we have effectively arrived at a specific form of
Example~1 of the Introduction.

Under the assumptions that have been made, it is reasonable, as part of Steps~1 and~3 of the
algorithm of Section~\ref{sec1}, to define the hypotheses $H_{P}$ and $H_{S}$ as follows:
\begin{gather}
H_{P}: \mu = 0 \nonumber\\
H_{S}: \rho(\hspace*{0.1em}\{\,X^* < - |x|\,\} \cup \{\,X^* > |x|\,\}\hspace*{0.1em})
= 2\Phi(- |x| / \sigma) \label{equ1}
\end{gather}
where $X^*$ is equal to an additional unobserved measurement of the concentration in question taken
at time $\mathtt{t}_2$ minus an additional unobserved measurement of the same type taken at time
$\mathtt{t}_1$, while $\Phi(y)$ is the cumulative density of a standard normal distribution at the
value $y$. It can be easily appreciated that these two hypotheses are in fact equiv\-alent. Observe
that the quantity on the right-hand side of the equality in equation~(\ref{equ1}) would be the
standard two-sided P value that would be calculated on the basis of the observation $x$ if $H_{P}$
was regarded as being the null hypothesis.

Now, in Step~4 of the algorithm of Section~\ref{sec1}, although a small value for this two-sided
P value would naturally disfavour the hypothesis $H_{S}$, and in particular favour the left-hand
side of the equality in equation~(\ref{equ1}) being greater than this P value, this would need to
be balanced by how much the pre-data assessment in Step 2 of this algorithm favoured the hypothesis
$H_{P}$.
Nevertheless, if the P value under discussion turns out to be very small then, even if the
hypothesis $H_{P}$ was quite strongly favoured before the data value $x$ was observed, it may well
be regarded as being rational to decide that hypothesis $H_{S}$ is fairly unlikely to be true.
As will always be the case, the evaluation of the likeliness of the hypothesis $H_{P}$ in Step~5 of
the algorithm in question should be the same as the evaluation of the likeliness of the hypothesis
$H_{S}$ in Step~4 of this algorithm.

\vspace*{3ex}
\subsection{Second example: Q values}
\label{sec23}

Let us now consider the more general case of the example being currently examined where the special
hypothesis in the scenario of Definition~2 is the almost sharp hypothesis that $\mu$ lies in the
interval $[-\varepsilon, \varepsilon]$, where $\varepsilon$ is a small positive constant. 

In this case, it is reasonable to define the hypotheses $H_{P}$ and $H_{S}$ as follows:
\begin{gather*}
H_{P}: \mu \in [- \varepsilon, \varepsilon]\\
H_{S}: \rho(\hspace*{0.1em}\{\,X^* < - |x|\,\} \cup \{\, X^* > |x|\,\}\hspace*{0.1em}) \leq
q(\varepsilon)
\end{gather*}
\par \vspace*{-1ex} \noindent
where
\vspace{-0.5ex}
\begin{equation}
\label{equ2}
q(\varepsilon) = \Phi((- |x| - \varepsilon) / \sigma) + \Phi((- |x| + \varepsilon) / \sigma)
\vspace*{1ex}
\end{equation}
It can easily be shown that these two hypotheses are equivalent under the assumptions that have
been made. Notice that the value $q(\varepsilon)$ as specified in equation~(\ref{equ2}) would be
classified as the Q value for the hypothesis that $\mu =\varepsilon$ according to the general
definition of a Q value that was presented and discussed in Bowater and Guzm\'an-Pantoja~(2019b),
and therefore here will also be referred to as a Q value.

Similar to the previous example, although we would naturally disfavour the hypothesis $H_{S}$, and
as a result, the hypothesis $H_{P}$ if this Q value was small, this would need to be balanced by
how much the hypothesis $H_{P}$ was favoured before the value $x$ was observed in order to make a
sensible evaluation of the likeliness of the hypothesis $H_{S}$ being true.

\vspace*{3ex}
\subsection{Third example: One-sided P values}
\label{sec4}

To give another example, let us look at an alternative way of defining the hypotheses $H_{P}$ and
$H_{S}$ in the context of the specific problem of inference that is currently under discussion. In
particular, let us now assume that, if $x \leq 0$, then these hypotheses would be defined as:
\vspace*{-1.5ex}
\begin{gather}
H_{P}: \mu \geq -\varepsilon \label{equ20}\\
H_{S}: \rho (X^* < x) \leq \Phi((x + \varepsilon) / \sigma) \label{equ21}
\end{gather}
while if $x > 0$, then they would have the definitions:
\vspace*{-0.5ex}
\begin{gather}
H_{P}: \mu \leq \varepsilon \label{equ22}\\
H_{S}: \rho (X^* > x) \leq \Phi((- x + \varepsilon) / \sigma) \label{equ4}
\end{gather}
Again, it can be easily shown that the hypotheses $H_{P}$ and $H_{S}$ in equations~(\ref{equ20})
and~(\ref{equ21}) are equivalent, and also that these hypotheses as defined in
equations~(\ref{equ22}) and~(\ref{equ4}) are equivalent.
In addition, observe that the quantities on the right-hand sides of the inequalities in
equations~(\ref{equ21}) and~(\ref{equ4}) would be the standard one-sided P values that would be
calculated on the basis of the observation $x$ if the null hypotheses were regarded as being the
hypotheses $H_{P}$ that correspond to the hypotheses $H_{S}$ defined in these two equations.

\pagebreak
Clearly, the substantial degree of pre-data belief that $\mu$ lies in the interval $[-\varepsilon,
\varepsilon]$ should be reflected in the pre-data assessment of the likeliness of the hypothesis
$H_{P}$ as defined in either equation~(\ref{equ20}) or equation~(\ref{equ22}).
Furthermore, similar to what was seen in the previous examples, a substantial degree of pre-data
belief in whichever one of the hypotheses $H_{P}$ in these equations is applicable would need to be
appropriately balanced by the information represented by the observation $x$ that is summarised by
the one-sided P value that appears in the corresponding hypothesis $H_{S}$, in order to make an
adequate assessment of the likeliness of this latter hypothesis given the observed value of $x$.

\vspace*{3ex}
\subsection{Discussion of examples}

Although the methods that have just been outlined in Sections~\ref{sec12} to~\ref{sec4} can be
applied to many other problems of inference than the simple one that has been considered, the
latter method based on one-sided P values is much more widely applicable than the former two
methods based on two-sided P values and on Q values, in particular, it is able to cope better with
sampling densities that are multimodal and/or non-symmetric.

Also, it can be argued, not just in terms of the specific problem that has been discussed but more
generally, that it is going to be less easy to evaluate, on the whole, the likeliness of hypotheses
$H_{S}$ that are based either on two-sided P values or on Q values than those that are based on
one-sided P values.
With regard to the examples that have been presented, a simple observation that underlies the
argument being referred to is that, for any given value of $x$, one of the two open intervals over
which either a two-sided P value or a Q value is determined by integration of the sampling density,
i.e.\ one of the intervals $(-\infty,-|x|)$ or $(|x|,\infty)$, will contain a proportion of the
sampling density that always decreases in size as the mean $\mu$ moves away from zero, despite of
course this change in $\mu$ always causing the total proportion of the sampling density contained
in these two intervals to increase.

For the reasons that have just been given, we will not consider generalising in a formal way the
methods based on two-sided P values and on Q values that were put forward in Sections~\ref{sec12}
and~\ref{sec23}. Instead, the type of method based on one-sided P values that was described in
Section~\ref{sec4} and developments of this method, will constitute the main form of bispatial
inference that will be explored in the rest of this paper.
It should be pointed out that, although it is apparent from the example considered in
Section~\ref{sec4} that possibly an important drawback of this method is that, in the scenario of
Definition~2, it will not generally allow us to directly assess the likeliness of the special
hypothesis that a parameter of interest $\nu$ lies in a narrow interval $[\nu_0, \nu_1]$ after the
data have been observed, i.e.\ that $\mu$ lies in the interval $[-\varepsilon,\varepsilon]$ in the
example in question, it will be shown later how this difficulty can be overcome.

\vspace*{3ex}
\section{Special form of bispatial inference}
\label{sec19}

\vspace*{1ex}
\subsection{General assumptions}
\label{sec21}

Let us now formalise the specific type of bispatial inference that has just been identified.

For the moment, it will be assumed that the only unknown parameter on which the sampling density
$g(x\,|\,\theta)$ depends is the parameter $\theta_j$, either because there are no other parameters
in the model, or because all the other parameters are known.
Also, we will assume that the scenario of interest is again the scenario outlined in Definition~2,
with the unknown parameter now being of course $\theta_j$, and that, in this scenario, the special
hypothesis is the almost sharp hypothesis that $\theta_j$ lies in the narrow interval
$[\theta_{j0},\theta_{j1}]$.

\vspace*{3ex}
\subsection{Test statistic}
\label{sec5}

Let us begin by detailing how the concept of a test statistic will be interpreted. In particular,
it will be assumed that a test statistic $T(x)$, which will also be denoted simply by the value
$t$, satisfies the following two requirements:

\vspace*{1.5ex}
\noindent
1) Similar to what in Bowater~(2019a) was defined as being a fiducial statistic, it is necessary
that the test statistic $T(x)$ is a univariate statistic of the sample $x$ that can be regarded as
efficiently summarising the information that is contained in this sample about the parameter
$\theta_j$, given the values of other statistics that do not provide any information about this
parameter, i.e.\ ancillary statistics.

\vspace*{1.5ex}
\noindent
2) Let $F(t \,|\, \theta_j, u)$ be the cumulative distribution function of the unobserved test
statistic $T(X)$ evaluated at its observed value $t$ given a value for the parameter $\theta_j$,
and conditional on $U(X)$ being equal to $u$, where $u$ are the observed values of an appropriate
set of ancillary statistics $U(X)$ of the data set of interest, i.e.\ $F(t \,|\, \theta_j, u)$
equals the probability $P(T(X) \leq t\,|\,\theta_j, u)$, and also let $\Fprime(t \,|\,\theta_j, u)
= P(T(X) \geq t\,|\,\theta_j, u)$.
On the basis of this notation, it is necessary that, over the set of allowable values for
$\theta_j$, the probabilities $F(t \,|\, \theta_j, u)$ and $1 - \Fprime(t \,|\, \theta_j, u)$
strictly decrease as $\theta_j$ increases.

\vspace*{1.5ex}
As far as the examples that will be considered in this paper are concerned, condition (1) will be
satisfied, in a simple and clear-cut manner, by $T(x)$ being a univariate sufficient statistic for
$\theta_j$. As a result, the set of ancillary statistics $U(x)$ referred to in condition (2) will
naturally be assigned to be empty in these examples, and in fact we could reasonably expect that
it would usually be appropriate to assign this set to be empty when the choice of the test
statistic $T(x)$ is more general.

\vspace*{3ex}
\subsection{Parameter and sampling space hypotheses}
\label{sec6}

If the condition
\begin{equation}
\label{equ3}
F(t \,|\, \theta_j=\theta_{j0}, u) \leq \Fprime(t \,|\, \theta_j=\theta_{j1}, u)
\vspace*{0.5ex}
\end{equation}
holds, where the values $\theta_{j0}$ and $\theta_{j1}$ are as defined in Section~\ref{sec21}, then
the hypotheses $H_{P}$ and $H_{S}$ will be defined as:\pagebreak
\begin{gather}
H_{P}: \theta_j \geq \theta_{j0} \label{equ23}\\
H_{S}: \rho( T(X^*) \leq t\,|\,u) \leq F(t\,|\,\theta_j = \theta_{j0}, u) \label{equ6}
\end{gather}
where $X^*$ is again an as-yet-unobserved sample of values drawn from the density function
$g(x\,|\,\theta)$ but now this sample will be assumed to be always of the same size as the observed
sample $x$, i.e.\ it must consist of $n$ observations, and where $\rho(T(X^*) \leq t\,|\,u)$ is the
\linebreak unknown population proportion of times that $T(X^*) \leq t$ conditional on the
ancillary \linebreak statistics $U(x)$ calculated on the basis of the data set $X^*$ being equal to
the values $u$, i.e.\ conditional on $U(X^*)=u$.
On the other hand, if the condition in equation~(\ref{equ3}) does not hold, then the hypotheses in
question will be defined as:
\begin{gather}
H_{P}: \theta_j \leq \theta_{j1} \label{equ24}\\
H_{S}: \rho( T(X^*) \geq t\,|\,u) \leq \Fprime(t\,|\,\theta_j = \theta_{j1}, u)
\label{equ7}
\end{gather}

\vspace*{0.5ex}
Given the way that the test statistic $T(x)$ was defined in Section~\ref{sec5}, it can be easily
appreciated that the hypotheses $H_{P}$ and $H_{S}$ in equations~(\ref{equ23}) and~(\ref{equ6}) are
equivalent, and also that these hypotheses as defined in equations~(\ref{equ24}) and~(\ref{equ7})
are equivalent.
In addition, observe that the probabilities $F(t\,|\,\theta_j = \theta_{j0},u)$ and
$\Fprime(t\,|\,\theta_j = \theta_{j1},u)$ that appear in the definitions of the hypotheses $H_{S}$
in equations~(\ref{equ6}) and~(\ref{equ7}) would be the standard one-sided P values that would be
calculated on the basis of the data set $x$ if the null hypotheses were regarded as being the
hypotheses $H_{P}$ that correspond to the two hypotheses $H_{S}$ in question.

We will assume that to make inferences about the parameter of interest $\theta_j$, the same
algorithm will be used as was outlined in Section~\ref{sec1}. However, with regard to the use of
this algorithm in the current context, let us make the following comments:

\vspace*{1ex}
\noindent
a) The set of parameters $\theta$ referred to in this algorithm will of course consist of only the
parameter $\theta_j$.

\pagebreak
\noindent
b) In Step~2 of this algorithm, it is evident that some special attention will often need to be
placed in assessing the likeliness of the almost sharp hypothesis that $\theta_j$ lies in the
interval $[\theta_{j0}, \theta_{j1}]$ based on only pre-data knowledge about $\theta_j$, since we
can see that this hypothesis will always be included in the hypothesis $H_{P}$, but will not
generally be equivalent to $H_{P}$.

\vspace*{1ex}
\noindent
c) In assessing the likeliness of the hypothesis $H_{S}$ in Step~4 of this algorithm, one of the
relevant factors that ought to be taken into account is clearly the size of the one-sided P value
that appears in the definition of this hypothesis, i.e.\ the value $F(t\,|\,\theta_j =
\theta_{j0},u)$ or the value $\Fprime(t\,|\,\theta_j = \theta_{j1},u)$.

\vspace*{1ex}
\noindent
d) Also in Step~4 of this algorithm, it now will be assumed that the goal is usually to assign a
probability to the hypothesis $H_{S}$.

\vspace*{1ex}
With reference to this last comment, the task of assigning a probability to the hypothesis $H_{S}$
may be made easier by first trying to determine what would be the minimum probability that could be
sensibly assigned to this hypothesis.
In particular, for a reason that should be obvious, it would not seem sensible to assign a
probability to the hypothesis $H_{S}$ that is less than the probability that would be assigned to
this hypothesis if nothing or very little had been known about the parameter $\theta_j$ before the
data were observed.
One way, but not as yet a widely accepted way, of making inferences about $\theta_j$ in this latter
type of situation is to use the fiducial method of inference (which has its origins in Fisher~1935
and Fisher~1956) and, given the interpretation of the concept of probability being relied on in the
present paper (see Section~\ref{sec3}), it would seem appropriate to consider applying the form of
this type of inference that has been called subjective, or more recently, organic fiducial
inference, see Bowater~(2017), Bowater~(2018a) and Bowater~(2019a).
In this regard, let $P_{f}(H_{S})$ denote the post-data or fiducial probability that would be
assigned to the hypothesis $H_{S}$ as a result of applying this latter method of fiducial inference
if there had been no or very little pre-data knowledge about $\theta_j$. Therefore, this value
$P_{f}(H_{S})$ can be considered as being a minimum value for the post-data probability of the
hypothesis $H_{S}$ being true in the genuine scenario of interest, i.e.\ the scenario of
Definition~2.
This method for placing a potentially useful lower limit on the probability of the hypothesis
$H_{S}$ will be illustrated as a feature of the examples that will be described in the next two
sections.

\vspace*{3ex}
\subsection{First example: Inference about a normal mean with variance known}
\label{sec7}

Let us return to the example that was discussed in Section~\ref{sec4}. We can see that this example
fits within the special framework for bispatial inference that has just been outlined.
In particular, the value $x$, i.e.\ the observed change in concentration, is clearly a suitable
test statistic $T(x)$, since it is a sufficient statistic for the mean $\mu$ that will satisfy
condition~(2) of Section~\ref{sec5} for any value it may possibly take.
Also, the way that the hypotheses $H_{P}$ and $H_{S}$ were specified in Section~\ref{sec4} matches
how these hypotheses would be specified by using the definitions in Section~\ref{sec6}.

In this earlier example, let us now more specifically assume that $\sigma=1$, $\varepsilon=0.2$ and
$x=2.7$. Under these assumptions, the relevant hypotheses $H_{P}$ and $H_{S}$ are as given in
equations~(\ref{equ22}) and~(\ref{equ4}), and the one-sided P value on the right-hand side of the
inequality in equation~(\ref{equ4}) is 0.0062. Since this P value is obviously small, but not very
small, if a substantial probability of around $0.3$ would have been placed on the hypothesis that
$\mu \in [-0.2, 0.2]$ before the value $x$ was observed, it would seem possible to justify a
probability in the range of say 0.03 to 0.08 being placed on the hypothesis
$H_{S}: \rho(X^* > 2.7) \leq 0.0062$ being true, and as a result, on the hypothesis
$H_{P}: \mu \leq 0.2$ being true after the value $x$ has been observed.

The probability that would be assigned to this hypothesis $H_{P}$ after the value $x$ has been
observed by applying the strong fiducial argument (see Bowater~2019a) as part of the method of
organic fiducial inference would be equal to 0.0062, i.e.\ the one-sided P value of interest.
We therefore can regard the probability $P_f(H_S)$ referred to in the last section as being equal
to 0.0062 in this example. Since the form of reasoning under discussion could be considered as
justifying this value of 0.0062 as being a minimum value for the probability of the hypothesis
$H_{S}$ in the genuine scenario of interest, it is therefore appropriate that the range of values
for this probability that has been proposed is well above this minimum value.

\vspace*{2.5ex}
\subsection{Second example: Inference about a binomial proportion}
\label{sec8}

Let us imagine that a random sample of patients are switched from being given a standard drug A to
being given a new drug B. After a period of time has passed, they are asked which out of the two
drugs A and B they prefer. The proportion of patients who prefer drug B to drug A, after patients
who do not express a preference have been excluded, will be denoted by the value $\mathtt{b}$.
Given this sample proportion, it will be assumed that the aim is make inferences about its
corresponding population proportion $\pi$.
For a similar reason with regard to the nature of drugs A and B as that given in Example 2 of the
Introduction, let us also suppose that the scenario of Definition~2 applies with the special
hypothesis being the hypothesis that the proportion $\pi$ lies in a narrow interval centred at 0.5,
which will be denoted as $[0.5-\varepsilon, 0.5+\varepsilon]$.

Observe that the sample proportion $\mathtt{b}$ clearly satisfies the requirements of
Section~\ref{sec5} to be a suitable test statistic $T(x)$. To give a more specific example, we will
assume that there are twelve patients in the sample, of whom nine prefer drug A to drug B, one
prefers drug B to drug A and two do not express a preference, and therefore $\mathtt{b}=0.1$.
Also, let the constant $\varepsilon$ be equal to 0.03. It now follows that, under the definitions
of Section~\ref{sec6}, the hypotheses $H_{P}$ and $H_{S}$ would be specified as:
\begin{gather}
H_{P}: \pi \geq 0.47 \nonumber\\
H_{S}: \rho( \mathtt{B}^* \leq 0.1) \leq 0.53^{10}+10(0.47)(0.53^{9}) = 0.0173 \label{equ5}
\end{gather}
where $\mathtt{B}^*$ is the proportion of patients who would prefer drug B to drug A in an
as-yet-unobserved sample of ten patients who express a preference between the two drugs.

We can see that again the one-sided P value, i.e.\ the value 0.0173 in equation~(\ref{equ5}), is
reasonably small. Therefore, if a pre-data probability of say 0.3 would have been placed on the
hypothesis that $\pi \in [0.47, 0.53]$, it would seem possible to justify a probability in the
range of say 0.03 to 0.08 being placed on the hypothesis $H_{S}$ being true, and as a result, on
the hypothesis $H_{P}$ being true after the proportion $\mathtt{b}$ has been observed.

The probability that would be assigned to the hypothesis $H_{P}$ after the value $\mathtt{b}$ has
been observed by using the strong fiducial argument, and a local pre-data (LPD) function for $\pi$
(see Bowater~2019a) defined by:
\vspace*{-0.5ex}
\begin{equation}
\label{equ16}
\omega_L (\pi) = \mathtt{c}\ \ \ \mbox{for all $\pi \in [0,1]$}
\vspace*{-0.5ex}
\end{equation}
where $\mathtt{c}$ is a positive constant, as part of the method of organic fiducial inference
would be equal to 0.0070.
Since this post-data probability can justifiably be regarded as the probability $P_f(H_S)$ referred
to Section~\ref{sec6}, and therefore as being a minimum value for the probability of the hypothesis
$H_{S}$ in the genuine scenario of interest, it is appropriate that, similar to the previous
example, the range of values for the probability of this hypothesis that has been proposed is well
above this minimum value.

\vspace*{3ex}
\subsection{Foundational basis of the theory}
\label{sec14}

In the examples considered in the previous section and Section~\ref{sec7}, it was inherently
assumed that the smaller the size of the one-sided P value that appears in the hypothesis $H_{S}$,
the less inclined we should be to believe that this hypothesis is true. However, what is the
foundational basis for this assumption? We will now try to offer some kind of answer to this
question.

It can be seen that the two versions of the hypothesis $H_{S}$ in equations~(\ref{equ6})
and~(\ref{equ7}) can both be represented as:
\vspace*{0.5ex}
\begin{equation}
\label{equ9}
H_{S} : \rho(A) \leq \beta
\vspace*{0.5ex}
\end{equation}
where $A$ is a given condition and $\beta$ is a given one-sided P value. Therefore, the population
proportion of times condition $A$ is satisfied will be less than or equal to $\beta$ if the
corresponding hypothesis $H_{P}$ is true, or in other words, if the parameter $\theta_j$ is
restricted in the way that is specified by this latter hypothesis.
However, we could also calculate a post-data probability for condition $A$ being satisfied without
placing restrictions on the parameter $\theta_j$ by using the fiducial argument. In particular, the
post-data probability in question would be defined as:
\vspace*{0.5ex}
\begin{equation}
\label{equ8}
P_f(A) = \int_{A} \int_{-\infty}^{\infty} g(X^*\,|\,\theta_j,u) f(\theta_j\,|\,x)
d\theta_j dX^*
\vspace*{1.5ex}
\end{equation}
where $f(\theta_j\,|\,x)$ is an appropriate fiducial density function for the parameter $\theta_j$.
To clarify, the outer integral in this equation is over unobserved data sets $X^*$ that satisfy
condition~$A$.

It will be helpful if we now look at a specific example, and so let us again consider the example
discussed in Sections~\ref{sec4} and~\ref{sec7}. In this case, the fiducial density of the
parameter of interest $\mu$, i.e.\ the density $f(\mu\,|\,x)$, obtained by using the strong
fiducial argument is defined by the expression $\mu \sim \mbox{N}(x, \sigma^2)$.
On the basis of this fiducial density for $\mu$, it is simple to show, by using
equation~(\ref{equ8}), how we obtain the result that $P_f(A)$ equals $0.5$ for any given observed
value of $x$, where as we know $A=\{X^* < x\}$ or $A=\linebreak\{X^*>x\}$, which of course is a
special result that in fact could have been derived by using more direct fiducial reasoning.
We could interpret this result as meaning that the probability that we should assign to condition
$A$ being true if we had known nothing or very little about $\mu$ before the value $x$ was observed
should be 0.5.

Taking into account this interpretation, if we were to \pagebreak propose assigning a large
probability to the hypothesis $H_{S}$ being true when the P value $\beta$ in equation~(\ref{equ9})
was quite small, then it would seem fair if we were asked how we can justify doing this given the
large difference between this P value and the probability $P_f(A)$.
To be able to give a satisfactory answer to this question, it is reasonable to argue that the only
situation we could be in would be one in which, before the value $x$ was observed, there had been
a high degree of belief that the hypothesis $H_{P}$ was true, which in the context of the scenario
of Definition~2, would mean a high degree of pre-data belief that $\mu$ lay in the interval
$[-\varepsilon, \varepsilon]$. In this situation, we could argue that assigning a large probability
to the hypothesis $H_{S}$ when the P value $\beta$ is quite small can be justified due to the
importance that is attached to the probability $P_f(A)$ as a benchmark or reference value being
greatly diminished as a result of our strong pre-data opinion about $\mu$.

Furthermore, let us suppose that a given probability of $\alpha_0$ would be assigned to the
hypothesis $H_{S}$ being true if the P value $\beta$ was equal to a given value $\beta_0$ that is
less than \linebreak say 0.05.
Now, if we imagine a scenario in which the value of $\beta$ is less than $\beta_0$, and therefore
further away from the probability $P_f(A)$ than $\beta_0$, then it can be easily argued that the
only way we could justify assigning the same probability $\alpha_0$ to the hypothesis $H_{S}$ would
be if, for an unrelated reason, it was decided that our pre-data belief that $\mu \in
[-\varepsilon, \varepsilon]$ should be increased. Also, it is fairly uncontroversial to argue that
for any \linebreak fixed data value $x$, the degree of pre-data belief that $\mu \in [-\varepsilon,
\varepsilon]$ and the degree of post-data belief in the hypothesis $H_{P}$ should be positively
correlated.
As a logical consequence of these arguments, it follows that, if there is a fixed degree of
pre-data belief that $\mu \in [-\varepsilon, \varepsilon]$, then the probability we should wish to
assign to the hypothesis $H_{S}$ after the value $x$ has been observed should decrease as the value
that the P value $\beta$ is assumed to take is made smaller, on condition that this P value is
already small.
Therefore, we hope that an adequate answer to the question posed at the start of this section has
been provided.

Another foundational issue that no doubt some would try to raise centres on the argument that the
probability that is assigned, on the basis of the observed data, to the hypothesis $H_{S}$ as part
of the method that has been outlined should be treated as a posterior probability that corresponds
to the Bayesian update of some given prior density function for the parameter of interest
$\theta_j$, where the choice for this prior density, of course, does not depend on the data.
However, to be able to sensibly use the Bayesian method being referred to some justification would
need to be given as to why such a prior density function would have been a good representation of
our beliefs about the parameter $\theta_j$ before the data were observed.
The fact that, in the context of the scenario of interest in Definition~2, it is going to be
extremely difficult, in general, to provide such a justification is consistent with the motivation
for the method of bispatial inference that was given in the Introduction.

\vspace*{3ex}
\section{Bispatial-fiducial inference}
\label{sec11}

The methodology of Sections~\ref{sec21} to~\ref{sec6} allows us to determine a post-data
probability for the hypothesis $H_{P}$ being true. Clearly though, it would be preferable to have a
post-data density function for the parameter of interest $\theta_j$. For this reason, let us now
consider generalising the methodology that has been proposed.

In particular, if
\vspace*{0.5ex}
\[
F(t \,|\, \theta_j=\theta_{j*}, u) \leq \Fprime(t \,|\, \theta_j=\theta_{j*} , u)
\vspace*{0.5ex}
\]
where $\theta_{j*}$ is any given value of $\theta_j$, then let us define the hypotheses $H_{P}$ and
$H_{S}$ as:
\vspace*{-0.5ex}
\begin{gather}
H_{P}: \theta_j \geq \theta_{j*}\nonumber\\
H_{S}: \rho( T(X^*) \leq t\,|\,u) \leq F(t\,|\,\theta_j = \theta_{j*}, u),\label{equ25}\\[-6ex]
\nonumber
\end{gather}
\par \vspace*{0.5ex} \noindent
otherwise, we will define these hypotheses as:
\pagebreak
\begin{gather}
H_{P}: \theta_j \leq \theta_{j*}\nonumber\\
H_{S}: \rho( T(X^*) \geq t\,|\,u) \leq \Fprime(t\,|\,\theta_j =
\theta_{j*}, u)\label{equ26}\\[-6ex]
\nonumber
\end{gather}
\par \vspace*{0.5ex} \noindent
We can observe that a post-data distribution function for $\theta_j$ could be constructed if, for
each value of $\theta_{j*}$ within the range of allowable values for $\theta_j$, we were able to
consistently evaluate the post-data probability of the hypothesis $H_{S}$ that is applicable as
defined by either equation~(\ref{equ25}) or~(\ref{equ26}).
Obviously, it would be a little awkward to do this by directly assessing the likeliness of the
hypothesis $H_{S}$ being true for all the values of $\theta_{j*}$ concerned, however no assumption
has been made regarding whether assessments of this type should be made directly or indirectly.

Therefore, we now will consider a strategy in which only one of these probability assessments is
made directly, while all the other assessments of this type that are required will in effect be
made indirectly by using again the method of organic fiducial inference.
The application of this general strategy will be referred to as bispatial-fiducial inference.

\vspace*{3ex}
\subsection{First proposed method}
\label{sec10}

Under the assumption that the hypothesis $H_{S}$ satisfies the more conventional definition of this
type of hypothesis given in Section~\ref{sec6}, let us assume that, after the data have been
observed, we directly weigh up, and then determine a value for the probability of this hypothesis
being true. This post-data probability will be denoted by the value $\alpha$, i.e.\
$P(H_{S})=\alpha$.

Furthermore, it will be assumed that the method of organic fiducial inference is used to derive a
fiducial density function for $\theta_j$ conditional on $\theta_j$ not lying in the interval
$[\theta_{j0}, \theta_{j1}]$.
In this approach to inference, the global pre-data (GPD) function $\omega_G (\nu)$ (see
Bowater~2019a) offers the principal, if not exclusive, means by which pre-data beliefs about a
parameter of interest $\nu$ can be expressed.
Given that it is being assumed that, under the condition that $\theta_j$ does not lie in the
interval $[\theta_{j0}, \theta_{j1}]$, nothing or very little would have been known about
$\theta_j$ before the data were observed, it is appropriate to use a neutral GPD function for
$\theta_j$ that has the following form:
\vspace*{0.5ex}
\begin{equation}
\label{equ13}
\omega_G (\theta_j) = \left\{
\begin{array}{ll}
0\ \ & \mbox{if $\theta_j \in [\theta_{j0}, \theta_{j1}]$}\\[1ex]
d & \mbox{otherwise}
\end{array}
\right.
\vspace*{1ex}
\end{equation}
where $d>0$. On the basis of this GPD function, the fiducial density function of $\theta_j$ that is
of interest can often be derived by applying what, in Bowater~(2019a), was referred to as the
moderate fiducial argument (when Principle 1 of this earlier paper can be applied).
Alternatively, in accordance with what was also advocated in Bowater~(2019a), this fiducial density
can be more generally defined, with respect to the same GPD function for $\theta_j$, by the
following expression:
\begin{equation}
\label{equ10}
f(\theta_j\,|\, \theta_j \notin [\theta_{j0}, \theta_{j1}], x) = \mathtt{C_{0}}
f_{S}(\theta_j\,|\, x)
\end{equation}
where $\mathtt{C_{0}}$ is a normalising constant, and $f_{S}(\theta_j\,|\,x)$ is a fiducial density
for $\theta_j$ derived by applying the strong fiducial argument (as part of what is required by
either Principle~1 or Principle~2 of Bowater~2019a) that would be regarded as being a suitable
fiducial density for $\theta_j$ in a general scenario where it is assumed that there was no or very
little pre-data knowledge about $\theta_j$ over all possible values of $\theta_j$.

Given the assumptions that have been made, if the condition in equation~(\ref{equ3}) holds, which
implies that $H_{P}$ is the hypothesis that $\theta_j \geq \theta_{j0}$, then it can be deduced
that the post-data probability of the event $\theta_j \in [\theta_{j0}, \theta_{j1}]$ is defined
by:
\vspace*{-0.25ex}
\begin{equation}
\label{equ11}
P(\theta_j \in [\theta_{j0}, \theta_{j1}]\,|\,x) = \alpha - \lambda (1 - \alpha)
\vspace*{-0.5ex}
\end{equation}
where the probability $\alpha$ is as defined at the start of this section, and $\lambda$ is given
by:
\vspace*{1ex}
\[
\lambda = \frac{P_{f}(\theta_j > \theta_{j1} \,|\, \theta_j \notin [\theta_{j0}, \theta_{j1}],x)}
{P_{f}(\theta_j < \theta_{j0} \,|\, \theta_j \notin [\theta_{j0}, \theta_{j1}],x)}
\vspace*{1.5ex}
\]
where $P_f(A\,|\,\theta_j \notin [\theta_{j0}, \theta_{j1}],x)$ denotes the fiducial probability
\pagebreak of the event $A$ conditional on $\theta_j \notin [\theta_{j0}, \theta_{j1}]$ that is the
result of integrating the fiducial density $f(\theta_j\,|\, \theta_j \notin [\theta_{j0},
\theta_{j1}], x)$ specified by equation~(\ref{equ10}) over those values of $\theta_j$ that satisfy
the condition $A$.
Under the condition in equation~(\ref{equ3}), it also follows that the post-data density function
of $\theta_j$, which will be denoted simply as $p(\theta_j \,|\,x)$, is defined over all of its
domain except for the interval $[\theta_{j0}, \theta_{j1}]$ by the expression:
\vspace*{1ex}
\begin{equation}
\label{equ12}
p(\theta_j \,|\,x) = \left\{
\begin{array}{ll}
(1 - \alpha) f(\theta_j\,|\,\{ \theta_j < \theta_{j0} \}, x)\ \ \ &
\mbox{if $\theta_j < \theta_{j0}$}
\vspace*{1.5ex}\\
\lambda (1 - \alpha) f(\theta_j\,|\,\{ \theta_j > \theta_{j1} \}, x) &
\mbox{if $\theta_j > \theta_{j1}$}
\end{array}
\right.
\vspace*{2ex}
\end{equation}
where $f(\theta_j\,|\,B, x)$ denotes the fiducial density $f(\theta_j\,|\, \theta_j \notin
[\theta_{j0}, \theta_{j1}], x)$ specified by equation~(\ref{equ10}) conditioned on the event $B$.
To clarify, the density $p(\theta_j \,|\,x)$ is being referred to as a \emph{post-data} density
because over the restricted space for $\theta_j$ in question it is defined on the basis of the
post-data probability of the hypothesis $H_{S}$, i.e.\ the value $\alpha$, and the fiducial density
$f(\theta_j\,|\, \theta_j \notin [\theta_{j0}, \theta_{j1}], x)$, which of course is a particular
type of post-data density.
While it has been assumed that the condition in equation~(\ref{equ3}) holds, it should be obvious,
on the basis of symmetry, how to modify the definitions in equations~(\ref{equ11})
and~(\ref{equ12}) in cases where this condition does not hold, i.e.\ when $H_{P}$ is the hypothesis
that $\theta_j \leq \theta_{j1}$.

Notice that the assignment of a probability $\alpha$ to the hypothesis $H_{S}$ that is greater than
or equal to the minimum value $P_f(H_{S})$ for this probability that was referred to at the end of
Section~\ref{sec6} is a sufficient (but not a necessary) requirement for ensuring that the
probability $P(\theta_j \in [\theta_{j0}, \theta_{j1}]\,|\,x)$ given in equation~(\ref{equ11}) is
not negative.
To clarify, the probability $P_f(H_{S})$ can be now more specifically expressed as:
\vspace*{1ex}
\begin{equation}
\label{equ27}
P_f(H_{S}) = \int_{\mbox{\footnotesize $\theta_{j0}$}}^{\mbox{\footnotesize $\infty$}}
f_{S}(\theta_j\,|\,x) d\theta_j\ \ \mbox{if $H_{P}$ is $\theta_j \geq \theta_{j0}$}\ \
\mbox{or}\ \ =\int_{\mbox{\footnotesize $-\infty$}}^{\mbox{\footnotesize $\theta_{j1}$}}
f_{S}(\theta_j\,|\,x) d\theta_j\ \ \mbox{if $H_{P}$ is $\theta_j \leq \theta_{j1}$}
\vspace*{0.5ex}
\end{equation}
where the fiducial density $f_{S}(\theta_j\,|\,x)$ is defined as it was immediately after
equation~(\ref{equ10}).

Furthermore observe that, if we are in the general case \pagebreak where $\theta_{j1} \neq
\theta_{j0}$, then although the definitions in equations~(\ref{equ11}) and~(\ref{equ12}) do not
fully specify the form taken by the post-data density function of $\theta_j$, this may not be a
great problem if the aim is to only derive post-data probability intervals for $\theta_j$, i.e.\
intervals in which there is a given post-data probability of finding the true value of $\theta_j$.
This is because the narrow interval $[\theta_{j0}, \theta_{j1}]$ over which this post-data density
of $\theta_j$ is undefined may often lie wholly inside or outside of the probability intervals of
the type in question that are of greatest interest.
On the other hand, it will of course often be indispensible to have a full rather than a partial
definition of the post-data density $p(\theta_j \,|\,x)$, e.g.\ for determining the post-data
expectations of general functions of $\theta_j$, and for simulating values from this kind of
density function.

One way around this problem is to simply complete the definition of the post-data density function
of $\theta_j$ by assuming that, when $\theta_j$ is conditioned to lie in the interval
$[\theta_{j0}, \theta_{j1}]$, it has a uniform density function over this interval.
Therefore, the full definition of this post-data density would consist of what is required both by
equation~(\ref{equ12}), and by the expression:
\vspace*{-1.5ex}
\[
p(\theta_j\,|\,x) = (\alpha - \lambda (1 - \alpha))/(\theta_{j1} - \theta_{j0})\ \ \
\mbox{if $\theta_j \in [\theta_{j0}, \theta_{j1}]$}
\]
Again, since by definition the interval $[\theta_{j0}, \theta_{j1}]$ is narrow, this simple
solution to the problem concerned may, in some circumstances, be considered adequate.

Nevertheless, it is a method that has two clear disadvantages. First, the post-data density
function of $\theta_j$ that it gives rise to will, in general, be discontinuous at the values
$\theta_{j0}$ and $\theta_{j1}$.
Second, the way in which the post-data density of $\theta_j$ conditional on the event $\theta_j \in
[\theta_{j0}, \theta_{j1}]$ is determined does not take into account our pre-data beliefs about
$\theta_j$, or the information contained in the data. Therefore, we will now try to enhance the
methodology that has been considered in the present section with the aim of addressing these two
issues.

\vspace*{3ex}
\subsection{A more sophisticated method}
\label{sec9}

The method that has just been outlined is based on determining a fiducial density for $\theta_j$
conditional on $\theta_j$ lying outside of the interval $[\theta_{j0}, \theta_{j1}]$ using the
neutral GPD function for $\theta_j$ given in equation~(\ref{equ13}). We now will attempt to
construct a fiducial density for $\theta_j$ conditional on $\theta_j$ lying inside this interval
using a more general type of GPD function for~$\theta_j$.

In particular, it will be assumed that this GPD function has the following form:
\vspace*{1ex}
\begin{equation}
\label{equ14}
\oline{\omega}_G (\theta_j) = \left\{
\begin{array}{ll}
1 + \tau h(\theta_j)\ \ & \mbox{if $\theta_j \in [\theta_{j0}, \theta_{j1}]$}
\vspace*{1ex}\\
0\ \ & \mbox{otherwise}
\end{array}
\right.
\vspace*{1ex}
\end{equation}
where $\tau \geq 0$ is a given constant and $h(\theta_j)$ is a continuous unimodal density function
on the interval $[\theta_{j0}, \theta_{j1}]$ that is equal to zero at the limits of this interval.
On the basis of this GPD function, the fiducial density of $\theta_j$ conditional on the event
$\theta_j \in [\theta_{j0}, \theta_{j1}]$ can often be derived by applying what, in
Bowater~(2019a), was referred to as the weak fiducial argument (see this earlier paper for further
details).
Alternatively, in accordance with what was also advocated in Bowater~(2019a), this fiducial density
can be more generally defined, with respect to the same GPD function for $\theta_j$, by the
following expression:
\begin{equation}
\label{equ15}
f(\theta_j\,|\,\theta_j \in [\theta_{j0}, \theta_{j1}], x)
= \mathtt{C_{1}}\hspace*{0.1em} \oline{\omega}_G (\theta_j) f_{S}(\theta_j\,|\,x)
\end{equation}
where the fiducial density $f_{S}(\theta_j\,|\,x)$ is specified as it was immediately after
equation~(\ref{equ10}) and $\mathtt{C_{1}}$ is a normalising constant.
Let us therefore make the assumption that this expression is used in conjunction with expressions
identical or analogous to the ones given in equations~(\ref{equ11}) and~(\ref{equ12}) in order to
obtain a full definition of the post-data density function of $\theta_j$ over all values of
$\theta_j$.

More specifically, though, it will be assumed that the constant $\tau$ in equation~(\ref{equ14}) is
chosen such that this overall density function for $\theta_j$ is made equivalent to a fiducial
density function for $\theta_j$ that is based on a continuous GPD function for $\theta_j$ over all
values of $\theta_j$. However, we will suppose that, except for the way in which the GPD function
of $\theta_j$ is specified, this fiducial density is derived on the basis of the same assumptions
as were used to derive the fiducial density $f_{S}(\theta_j\,|\,x)$.
If the hypothesis $H_{S}$ is assigned a probability $\alpha$ that is greater than the minimum value
for this probability discussed in Section~\ref{sec6}, i.e.\ the probability $P_f(H_{S})$ defined by
equation~(\ref{equ27}), and the density $f_{S}(\theta_j\,|\,x)$ is positive for all allowable
values of $\theta_j$, then the value of $\tau$ in question will always exist and be unique.

This criterion for choosing the value of $\tau$ will, in general, ensure that the post-data density
function for $\theta_j$ being considered will be continuous over all values of $\theta_j$. Also, if
$\theta_j$ was conditioned to lie in the interval $[\theta_{j0}, \theta_{j1}]$, then this post-data
density would still be formed in a way that takes into account our pre-data beliefs about
$\theta_j$, and then allows these beliefs to be modified on the basis of the data.
Therefore, the difficulties are avoided that were identified as being associated with the method
that was proposed in the previous section.

Furthermore, there are two reasons why the criterion in question concerning the choice of the
constant $\tau$ can be viewed as not being a substantial restriction on the way we are allowed to
express our pre-data knowledge about the parameter $\theta_j$. First, since going against this rule
will in general lead to the post-data density of $\theta_j$ having undesirable discontinuities, it
can be regarded as being a useful guideline in choosing a suitable GPD function for $\theta_j$ when
$\theta_j$ is conditioned to lie in the interval $[\theta_{j0}, \theta_{j1}]$.
Second, any detrimental effect caused by enforcing this criterion may not be that apparent given
the great deal of imprecision there will usually be in the specification of this GPD function.

\vspace*{3ex}
\subsection{First example: Revisiting the simple normal case}
\label{sec15}

To give an example of the application of the method proposed in the previous section, let us return
to the problem of making inferences about the mean $\mu$ of a normal distribution that was
considered in both Sections~\ref{sec4} and~\ref{sec7}. All assumptions about the values of
quantities of interest that were made in Section~\ref{sec7} will be maintained.

In this example, the fiducial density $f(\theta_j\,|\, \theta_j \notin [\theta_{j0},
\theta_{j1}], x)$ that was defined in Section~\ref{sec10}, which now of course can be denoted as
the density $f(\mu\,|\,\mu \notin [-0.2, 0.2], x)$, can be directly derived by using the moderate
fiducial argument (see Bowater~2019a), which implies that it is simply the normal density function
of $\mu$ defined by the expression $\mu \sim \mbox{N} (2.7, 1)$ conditioned not to lie in the
interval $[-0.2, 0.2]$.
To clarify, using the more general definition of the density $f(\theta_j\,|\, \theta_j \notin
[\theta_{j0}, \theta_{j1}], x)$ in equation~(\ref{equ10}), the fiducial density
$f_{S}(\theta_j\,|\,x)$, i.e.\ the density $f_{S}(\mu\,|\,x)$ in the present case, would be the
normal density of $\mu$ just mentioned.

Let us now make the assumption that the density function $h(\theta_j)$ that appears in
equation~(\ref{equ14}) is defined as being a beta density function for $\mu$ on the interval
$[-0.2,0.2]$ with both its shape parameters equal to 4. This density function clearly satisfies the
conditions on the function $h(\theta_j)$ that were given in Section~\ref{sec9}. Furthermore, it is
a reasonable choice for this function given that it is smooth, its mode is at $\mu=0$ and it is
symmetric.

Given this assumption, if a sensible value $\alpha$ was assigned to the probability of the
hypothesis $H_{S}$, then the precise form of the post-data density $p(\mu\,|\,x)$ would be
determined by the expression in equation~(\ref{equ15}) and expressions analogous to the ones given
in equations~(\ref{equ11}) and~(\ref{equ12}).
On the other hand, of course, this density function could have been defined according to the simple
proposals for both its partial and full specification outlined in Section~\ref{sec10} without the
need to have made an assumption about the form of the density $h(\theta_j)$.

The curves in Figure~1 represent the post-data density $p(\mu\,|\,x)$ outside of the interval
$[-0.2,0.2]$ for all the definitions of this density function being considered, while, over the
whole of the real line, they more specifically represent this function under its more sophisticated
definition given in Section~\ref{sec9}. The range of values for the probability $P(H_{S})$, i.e.\
the probability $\alpha$, that underlies this figure is equal to what was proposed as being
appropriate for this example in Section~\ref{sec7}, i.e.\ the range of 0.03 to 0.08.
In particular, the solid curve in Figure~1 depicts the post-data density $p(\mu\,|\,x)$ when
$\alpha$ is 0.05, while the long-dashed and short-dashed curves in this figure depict this density
when $\alpha$ is 0.03 and 0.08 respectively.
The accumulation of probability mass around the value of zero in these density functions is
consistent with what we would have expected given the strong pre-data belief that $\mu$ would be
close to zero, though as we know, the importance of this pre-data opinion about $\mu$ is assessed
in the context of having observed the data value $x$ to obtain the density functions that are
shown.

\begin{figure}[t]
\begin{center}
\includegraphics[width=5in]{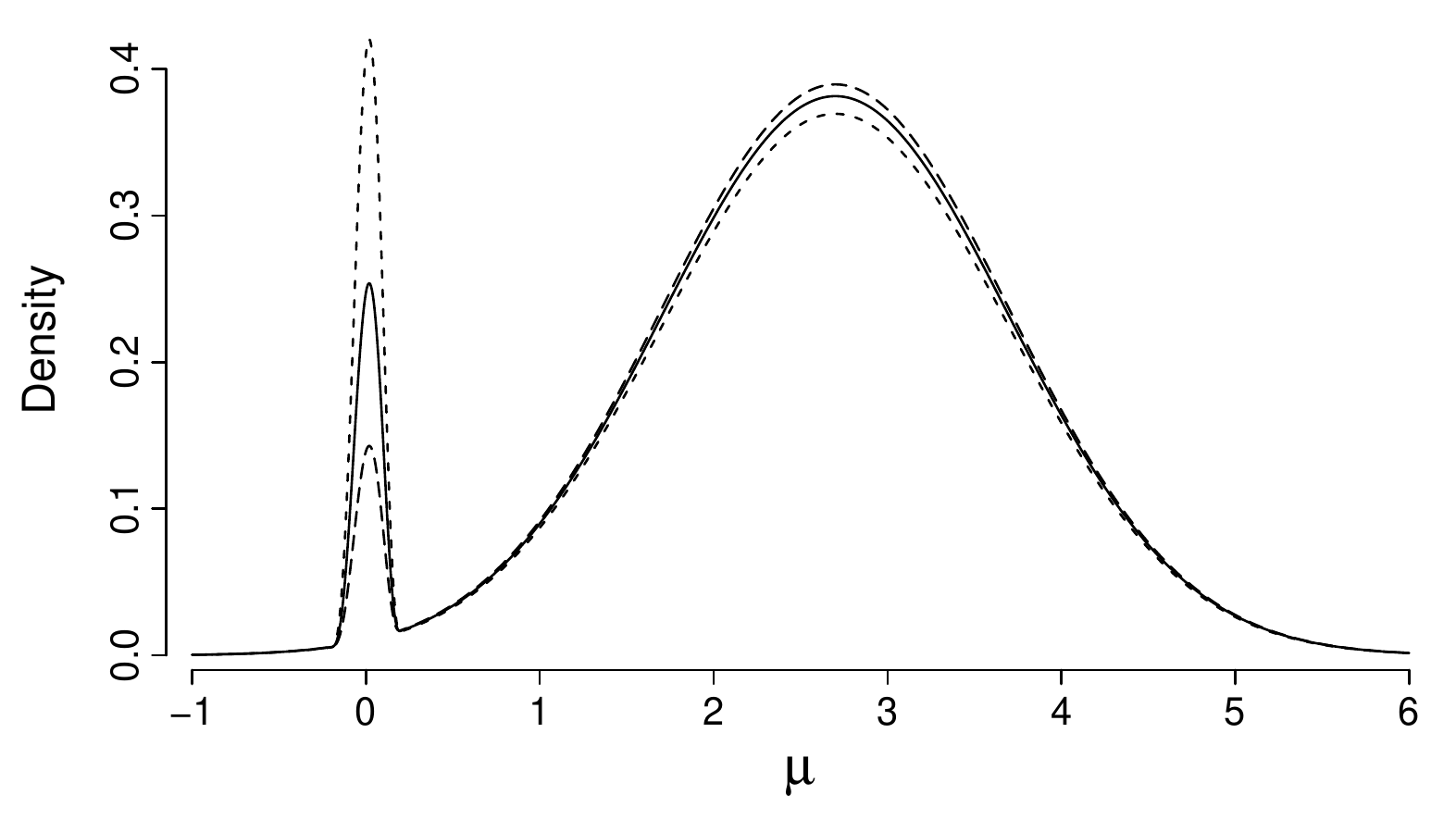}
\caption{\small{Post-data densities of a normal mean $\mu$ when $\sigma^2$ is known}}
\end{center}
\end{figure}

\vspace*{3ex}
\subsection{Second example: Revisiting the binomial case}
\label{sec16}

To give a second example of the application of the method being discussed, let us consider once
more the problem of making inferences about a binomial proportion $\pi$ that was examined in
Section~\ref{sec8}, with the same assumptions in place about the values of quantities of interest
that were made in this earlier section.

In using the method of organic fiducial inference in this example to derive the fiducial density
$f_{S}(\pi\,|\,x)$ that was specified for the general case immediately after equation~(\ref{equ10})
and which is required as an input in both this earlier equation and equation~(\ref{equ15}), it
would again seem reasonable to define the LPD function $\omega_L(\pi)$ as in
equation~(\ref{equ16}).
Also, as explained in Bowater~(2019a), this fiducial density for $\pi$ will be fairly insensitive
to the choice made for the LPD function in question.
We will, in addition, assume that the density $h(\theta_j)$ required by equation~(\ref{equ14}) is a
beta density with both its shape parameters equal to 4, as it was in the previous example, but this
time defined on the interval $[0.47, 0.53]$.
As a result, we now can specify the post-data density $p(\pi\,|\,x)$ using equations~(\ref{equ11}),
(\ref{equ12}) and~(\ref{equ15}), assuming again, of course, that a sensible value $\alpha$ has been
assigned to the probability of the hypothesis $H_{S}$.

The histogram in Figure~2, which strictly speaking is a weighted histogram, represents this
post-data density for the case where $\alpha=0.05$.
The numerical output on which this histogram is based was generated by the method of importance
sampling, more specifically by appropriately weighting a sample of three million independent random
values from the fiducial density $f_{S}(\pi\,|\,x)$.
On the other hand, the curves in Figure~2 represent approximations to the post-data density
$p(\pi\,|\,x)$ obtained by substituting the fiducial density $f_{S}(\pi\,|\,x)$ used in its
construction by the posterior density of the proportion $\pi$ that is based on the Jeffreys prior
density for this problem, or in other words, based on choosing the prior density of $\pi$ to be a
beta density function of $\pi$ with both its shape parameters equal to 0.5.
Additional simulations showed that this method of approximation was satisfactory.
In particular, the solid curve in Figure~2 approximates the density $p(\pi|\,x)$ for the case where
$\alpha=0.05$ and, as was expected, it closely approximates the histogram in this figure.
Similarly, the long-dashed and short-dashed curves in this figure approximate the density in
question in the cases where $\alpha$ is 0.03 and 0.08 respectively.
Again, the range of values for $\alpha$ being considered is 0.03 to 0.08, which is equal to what
was proposed as being appropriate for this example in Section~\ref{sec8}.

\begin{figure}[t]
\begin{center}
\includegraphics[width=5in]{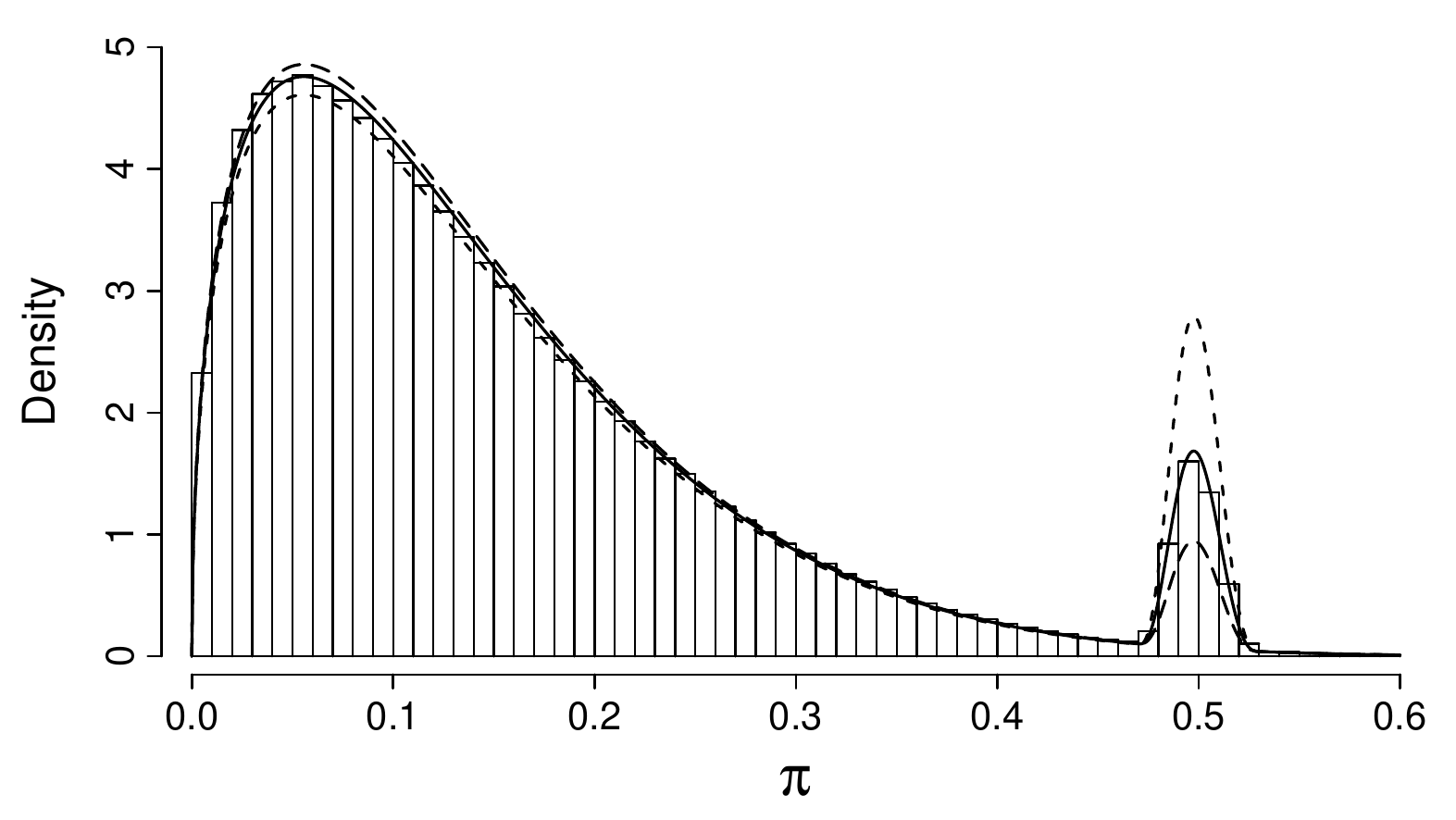}
\caption{\small{Post-data densities of a binomial proportion $\pi$}}
\end{center}
\end{figure}

The accumulation of probability mass around the value of 0.5 in the density functions in Figure~2
is clearly an artefact of the strong pre-data opinion that was held about the proportion $\pi$.

\vspace*{3ex}
\section{Extending bispatial-fiducial inference to multi-parameter problems}
\label{sec20}

\vspace*{1ex}
\subsection{General assumptions}
\label{sec13}

It was assumed in Section~\ref{sec24} that the parameter $\theta_j$ is the only unknown parameter
in the sampling model. Let us now assume that all the parameters $\theta_1, \theta_2, \ldots,
\theta_k$ on which the sampling density $g(x\,|\,\theta)$ depends are unknown.

More specifically, we will assume that the subset of parameters $\theta^{A} = \{ \theta_1,
\theta_2, \ldots, \theta_m \}$ is such that what would have been believed, before the data were
observed, about each parameter $\theta_{j}$ in this set if all other parameters in the model, i.e.\
$\theta_{-j} = \{ \theta_1, \ldots, \linebreak \theta_{j-1},$ $\theta_{j+1},\ldots,\theta_{k} \}$,
had been known, would have satisfied the requirements of the scenario of Definition~2 with
$\theta_j$ being the parameter of interest $\nu$.
Furthermore, it will be assumed that the set of all the remaining parameters in the sampling model,
i.e.\ the set $\theta^{B} = \{ \theta_{m+1}, \theta_{m+2}, \ldots, \theta_k \}$, is such that,
before the data were observed, nothing or very little would have been known about each parameter
$\theta_{j}$ in this set over all allowable \linebreak values of the parameter if all other
parameters in the model, i.e.\ the set $\theta_{-j}$, had been known.

To derive the post-data density of any given parameter $\theta_j$ in the set $\theta^{A}$
conditional on all the parameters in the set $\theta_{-j}$ being known, we can clearly justify
applying the method outlined in Section~\ref{sec9}, meaning that this density function would be
defined by equation~(\ref{equ15}) and by expressions identical or analogous to the ones given in
equations~(\ref{equ11}) and~(\ref{equ12}).
The set of full conditional post-data densities that result from doing this for each parameter in
the set $\theta^{A}$ would therefore be naturally denoted as:
\begin{equation}
\label{equ17}
p(\theta_j\,|\,\theta_{-j},x)\ \ \ \ j=1,2,\ldots,m
\end{equation}
It is also clearly defensible to specify the post-data density of any given parameter $\theta_j$ in
the set $\theta^{B}$ conditional on all the parameters in the set $\theta_{-j}$ being known as
being the fiducial density $f_{S}(\theta_j\,|\,x)$ that was defined immediately after
equation~(\ref{equ10}).
Doing this for each parameter in the set $\theta^{B}$ would therefore give rise to the following
set of full conditional post-data densities:
\begin{equation}
\label{equ18}
f_{S}(\theta_j\,|\,\theta_{-j},x)\ \ \ \
j=m\hspace*{-0.05em}+\hspace*{-0.05em}1,m\hspace*{-0.05em}+\hspace*{-0.05em}2,\ldots,k
\end{equation}

\vspace*{3ex}
\subsection{Post-data densities of various parameters}
\label{sec22}

If the complete set of full conditional densities of the parameters $\theta_1, \theta_2, \ldots,
\theta_k$ that results from combining the sets of full conditional densities in
equations~(\ref{equ17}) and~(\ref{equ18}) determine a unique joint density for these parameters,
then this density function will be defined as being the post-data density function of $\theta_1,
\theta_2, \ldots, \theta_k$ and will be denoted as $p(\theta\,|\,x)$.
However, this complete set of full conditional densities may not be consistent with any joint
density of the parameters concerned, i.e.\ these full conditional densities may be incompatible
among themselves.

To check whether the full conditional densities of $\theta_1, \theta_2, \ldots, \theta_k$ being
referred to are compatible, it may be possible to use an analytical method. An example of an
analytical method that could be used to try to achieve this goal was outlined in relation to full
conditional densities of a similar type in Bowater~(2018a).

By contrast, in situations that will undoubtedly often arise where it is not easy to establish
whether or not the full conditional densities defined by equations~(\ref{equ17}) and~(\ref{equ18})
are compatible, let us imagine that we make the pessimistic assumption that they are in fact
incompatible.
Nevertheless, even though these full conditional densities could be incompatible, they could be
reasonably assumed to represent the best information that is available for constructing a joint
density function for the parameters $\theta$ that most accurately represents what is known about
these parameters after the data have been observed, i.e.\ constructing, what could be referred to
as, the most suitable post-data density for these parameters.
Therefore, it would seem appropriate to try to find the joint density of the parameters $\theta$
that has full conditional densities that most closely approximate those given in
equations~(\ref{equ17}) and~(\ref{equ18}).

To achieve this objective, let us focus attention on the use of a method that was advocated in a
similar context in Bowater~(2018a), in particular the method that simply consists in making the
assumption that the joint density of the parameters $\theta$ that most closely corresponds to the
full conditional densities in equations~(\ref{equ17}) and~(\ref{equ18}) is equal to the limiting
density function of a Gibbs sampling algorithm (Geman and Geman~1984, Gelfand and Smith~1990) that
is based on these conditional densities with some given fixed or random scanning order of the
parameters in question.
Under a fixed scanning order of the model parameters, we will define a single transition of this
type of algorithm as being one that results from randomly drawing a value (only once) from each of
the full conditional densities in equations~(\ref{equ17}) and~(\ref{equ18}) according to some given
fixed ordering of these densities, replacing each time the previous value of the parameter
concerned by the value that is generated.
Let us clarify that it is being assumed that only the set of values for the parameters $\theta$
that are obtained on completing a transition of this kind are recorded as being a newly generated
sample, i.e.\ the intermediate sets of parameter values that are used in the process of making such
a transition do not form part of the output of the algorithm.

To measure how close the full conditional densities of the limiting density function of the general
type of Gibbs sampler being presently considered are to the full conditional densities in
equations~(\ref{equ17}) and~(\ref{equ18}), we can make use of a method that, in relation to its use
in a similar context, was discussed in Bowater~(2018a).
To be able to put this method into effect it is required that the Gibbs sampling algorithm that is
based on the full conditional densities in equations~(\ref{equ17}) and~(\ref{equ18}) would be
irreducible, aperiodic and positive recurrent under all possible fixed scanning orders of the
parameters $\theta$. \linebreak
Assuming that this condition holds, it was explained in Bowater~(2018a), how it may be useful to
analyse how the limiting density function of the Gibbs sampler in question varies over a reasonable
number of very distinct fixed scanning orders of the parameters concerned, remembering that each of
these scanning orders has to be implemented in the way that was just specified.
In particular, it was concluded that if within such an analysis, the variation of this limiting
density with respect to the scanning order of the parameters $\theta$ can be classified as small,
negligible or undetectable, then this should give us reassurance that the full conditional
densities in equations~(\ref{equ17}) and~(\ref{equ18}) are, respectively according to such
classifications, close, very close or at least very close, to the full conditional densities of the
limiting density of a Gibbs sampler of the type that is of main interest, i.e.\ a Gibbs sampler
that is based on any given fixed or random scanning order of the parameters concerned.
See Bowater~(2018a) for the line of reasoning that justifies this conclusion.

In trying to choose the scanning order of the parameters $\theta$ such that the type of Gibbs
sampler under discussion has a limiting density function that corresponds to a set of full
conditional densities that most accurately approximate the density functions in
equations~(\ref{equ17}) and~(\ref{equ18}), we should always take into account the precise context
of the problem of inference being analysed.
Nevertheless, a good general choice for this scanning order could arguably be, what will be
referred to as, a uniform random scanning order. Under this type of scanning order, a transition of
the Gibbs sampling algorithm in question will be defined as being one that results from generating
a value from one of the full conditional densities in equations~(\ref{equ17}) and~(\ref{equ18})
that is chosen at random, with \linebreak the same probability of $1/k$ being given to any one of
these densities being selected, \linebreak and then treating the generated value as the updated
value of the parameter concerned.

It is clear that being able to obtain a large random sample from a suitable post-data density of
the parameters $\theta_1, \theta_2, \ldots, \theta_k$ using a Gibbs sampler in the way that has
been described in the present section will usually allow us to obtain good approximations to
expected values of given functions of these parameters over the post-data density concerned, and
thereby allow us to make useful and sensible inferences about the parameters $\theta_1, \theta_2,
\ldots, \theta_k$ on the basis of the data set $x$ of interest.

\vspace*{3ex}
\subsection{Post-data opinion curve}

In constructing any one of the post-data densities $p(\theta_j\,|\,\theta_{-j},x)$ in
equation~(\ref{equ17}), there is, nevertheless, still one important issue that needs to be
addressed, which is that the assessment of the likeliness of the relevant hypothesis $H_{S}$ in
equation~(\ref{equ6}) or~(\ref{equ7}) will generally depend on the values of the parameters in the
set $\theta_{-j}$.
This of course will be partially due to the effect that the values of these parameters can have on
the one-sided P value that appears in the definition of this hypothesis, i.e.\ the P value $\beta$
in terms of the notation used in equation~(\ref{equ9}).
Therefore, in general, we will not need to assign just one probability to the hypothesis $H_{S}$,
but various probabilities conditional on the values of the parameters in the set $\theta_{-j}$.

Faced with the inconvenience that this can cause, it is possible to simplify matters greatly by
assuming that the probability that is assigned to any given hypothesis $H_{S}$, i.e.\ the
probability $\alpha$, will be the same for any fixed value of the one-sided P value $\beta$ that
appears in the definition of this hypothesis, no matter what values are actually taken by the
parameters in the set $\theta_{-j}$.
It can be argued that such an assumption would be reasonable in many practical situations.
If this assumption is made then, the probability $\alpha$ will clearly be a mathematical function
of the one-sided P value $\beta$.
We will call this function the post-data opinion (PDO) curve for the parameter $\theta_j$
conditional on the parameters $\theta_{-j}$.

Notice that, when using the Gibbs sampling method outlined in the last section, we will only need
to define this curve over the range of values of the P value $\beta$ that are likely to appear in
the hypothesis $H_{S}$ having taken into account the data $x$ that have been observed.
Also, it could be hoped that, in many cases, it would be possible to adequately specify this curve
by first assessing the probability $\alpha$ for a small number of carefully selected values of
$\beta$, and then fitting some type of smooth curve through the resulting points.

Furthermore, it is clear that there are rules which can be employed to ensure that any given PDO
curve is chosen in a sensible manner. Perhaps the most obvious rule of this kind is that a PDO
curve, in general, must be chosen such that, along this curve, the value of $\alpha$ monotonically
increases as the P value $\beta$ is increased from zero up to its maximum permitted value in the
particular case of interest.
Other characteristics that we would expect this type of curve to have will be discussed in the
context of the examples that will be considered in the next two sections.

\vspace*{3ex}
\subsection{First example: Normal case with variance unknown}
\label{sec17}

To give an example of the application of the method of inference that has just been proposed, i.e.\
bispatial-fiducial inference for multi-parameter problems, let us return to the example that was
first considered in Section~\ref{sec8}, however let us now assume that the difference in the
performance of the two drugs of interest is measured by the difference in the concentration of a
certain chemical (e.g.\ cholesterol) in the blood, in particular the level observed for drug A
minus the level observed for drug B, rather than the prefer\-ences of the patients concerned.
The set of these differences for all of the $n$ patients in the sample will be the data set $x$.
This example therefore also shares notable common ground with the example that was first considered
in Section~\ref{sec12}. Moreover, similar to this earlier example, it will be assumed that each
value $x_i$ in the data set $x$ follows a normal distribution with an unknown mean $\mu$, however,
by contrast to this previous example, the standard deviation $\sigma$ of this distribution will now
also be assumed to be unknown.

For the same type of reason to that used in Example~2 in the Introduction, let us in addition
suppose that, for any given value of $\sigma$, the scenario of Definition~2 would apply in relation
to the parameter $\mu$ with the special hypothesis being the hypothesis that $\mu$ lies in the
narrow interval $[-\varepsilon, \varepsilon]$.
On the other hand, it will be assumed, as could often be done in practice, that nothing or very
little would have been known about the standard deviation $\sigma$ before the data were observed
given any value for $\mu$. Therefore, in terms of the notation of Section~\ref{sec13}, the set of
parameters $\theta^{A}$ will only contain $\mu$, and the set $\theta^{B}$ will only
contain~$\sigma$.

To determine the post-data density $p(\mu\,|\,\sigma,x)$, the test statistic $T(x)$ as defined in
Section~\ref{sec5} will be assumed to be the sample mean $\bar{x}$, which clearly satisfies what is
required to be such a statistic. Under this assumption, the hypotheses $H_{P}$ and $H_{S}$ will be
as given in Section~\ref{sec4}, except that now the mean $\bar{x}$ takes the place of the value
$x$, the standard error\hspace*{0.1em} $\sigma/\sqrt{n}$\hspace*{0.1em} takes the place of
$\sigma$, and the random variable $X^{*}$ is substituted by $\overbar{X}^{*}$, i.e.\ by the mean of
an as-yet-unobserved sample of $n$ additional observations from \linebreak the population
concerned.
If we more specifically assume, as we will do from now on, that $n=9$, $\bar{x}=2.7$ and
$\varepsilon=0.2$, then it is evident that, since the condition in equation~(\ref{equ3}) does not
hold in this case, the relevant hypotheses $H_{P}$ and $H_{S}$ can be defined as:
\begin{gather*}
H_{P}: \mu \leq 0.2\\
H_{S}: \rho (\overbar{X}^{*} > 2.7) \leq \Phi(- 7.5 / \sigma)\ \ (=\beta)
\end{gather*}
\par \vspace*{0.5ex}
Clearly, the one-sided P value $\beta$ in the hypothesis $H_{S}$ depends on the standard deviation
$\sigma$. Moreover, given that there is a one-to-one relationship between $\sigma$ and this P
value, the PDO curve for $\mu$ conditional on $\sigma$ will completely describe the assessment of
the probability of $H_{S}$, i.e.\ the probability $\alpha$, in all possible circumstances of
interest. To give a simple example, we will assume that this PDO curve has the algebraic form:
$\alpha = \beta^{\hspace*{0.1em}0.6}$. In Figure~3, this PDO curve is represented by the solid
curve.

\begin{figure}[t]
\begin{center}
\includegraphics[width=4in]{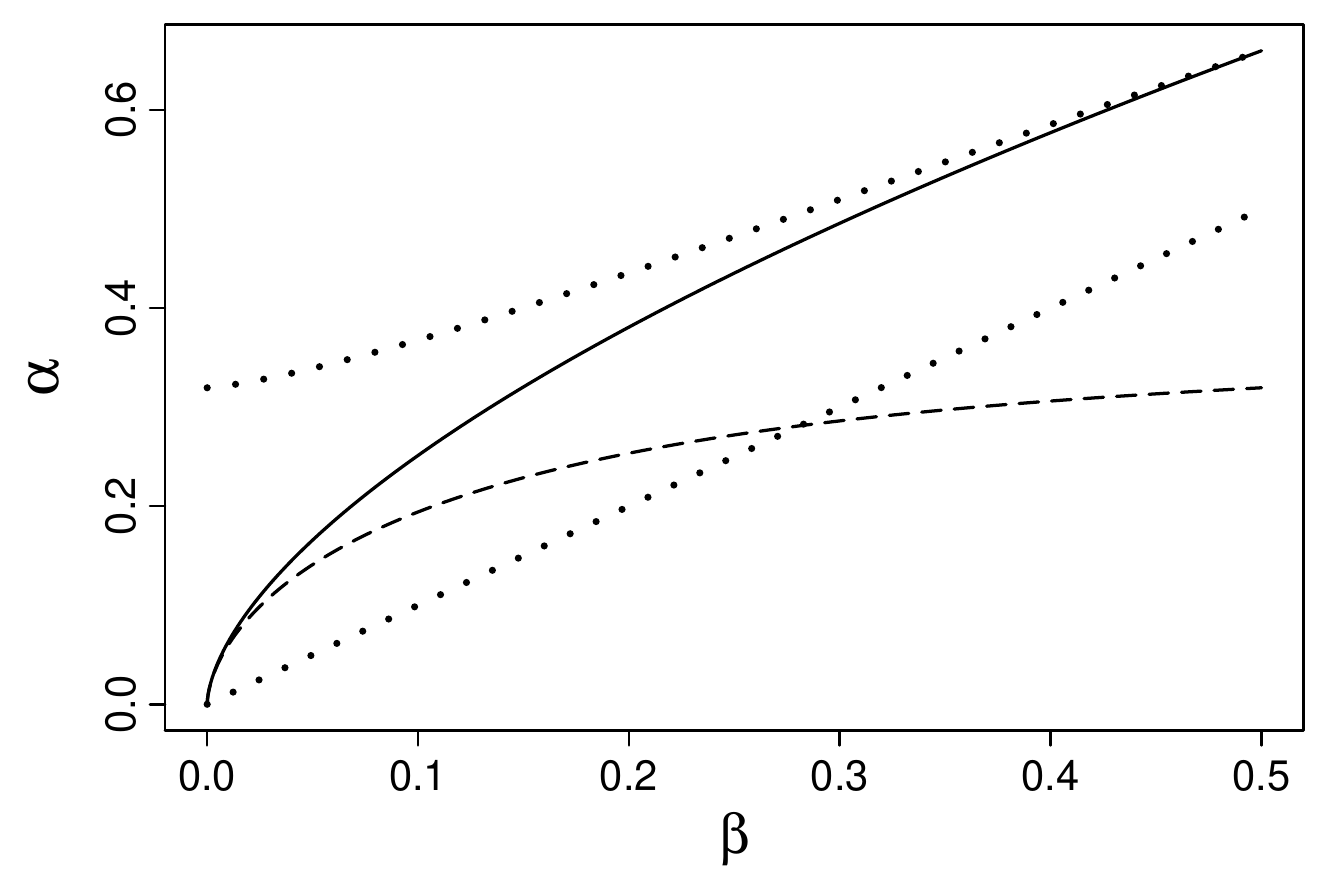}
\caption{\small{Post-data opinion (PDO) curve for $\mu$ conditional on $\sigma$}}
\end{center}
\end{figure}

Similar to the example that was considered in Section~\ref{sec14}, the fiducial density function of
$\mu$ conditional on $\sigma$ that is obtained by using the strong fiducial argument, i.e.\ the
density $f_{S}(\mu\,|\,\sigma, x)$, is defined by the expression
$\mu \sim \mbox{N}(2.7, \sigma^2/9)$. To clarify, what was referred to in Bowater~(2019a) as being
the fiducial statistic is here being assumed to be any sufficient statistic for $\mu$, e.g.\ the
sample mean $\bar{x}$. Using this simple analytical result, the lower dotted line in Figure~3
represents what the PDO curve under discussion would need to be so that the probability $\alpha$
equals the probability that would be assigned to the hypothesis $H_{S}$ by the fiducial density
$f_{S}(\mu\,|\,\sigma, x)$.
For the same reason that was given in Section~\ref{sec6}, we would expect any choice for the PDO
curve in question to be always higher than this dotted line, which is the case, as can be seen, for
the PDO curve that has been proposed.

Using this proposed PDO curve, i.e.\ the function $\alpha = \beta^{\hspace*{0.1em}0.6}$, and the
fiducial density $f(\mu\,|\,\mu \notin [-0.2,0.2],\sigma, x)$ defined by equation~(\ref{equ10}) as
inputs into the method described in Section~\ref{sec10} enables us to calculate, via
equation~(\ref{equ11}), the post-data probability conditional on $\sigma$ that $\mu$ lies in the
interval $[-0.2, 0.2]$, i.e.\ the probability $P(\mu \in [-0.2, 0.2]\,|\,\sigma,x)$, for any given
value of the P value $\beta$. The dashed curve in Figure~3 was constructed by plotting this
post-data probability $P(\mu \in [-0.2, 0.2]\,|\,\sigma,x)$, rather than the probability $\alpha$,
against different values for the P value $\beta$. It can be seen that this curve is monotonically
increasing, which is a desirable outcome. If this had not been the case, then it could have been
quite reasonably concluded that the PDO curve on which this dashed curve is based does not
represent logically structured beliefs about $\mu$ conditional on $\sigma$ in the context of what
has been assumed to have been known about $\mu$ before the data were observed.

Finally, the upper dotted curve in Figure~3 represents what the PDO curve of interest would need to
be if, under the assumptions already made, we decided that independent of the size of the P value
$\beta$, we would place a post-data probability on $\mu$ lying in the interval $[-0.2, 0.2]$ that
was equal to the limiting value assigned to this probability by the dashed curve as $\beta$ tends
to 0.5, i.e.\ a probability of 0.32. Clearly, if this limiting value of the probability in question
is regarded as being appropriate, then any sensible PDO curve in the case being considered would
need to lie below this dotted curve and, as can be seen, the proposed PDO curve also satisfies this
constraint.

If in addition we specify the density function $h(\theta_j)$, which is required by
equation~(\ref{equ14}), i.e.\ the function $h(\mu)$ in the present case, to be the same as it was
in the example in \linebreak Section~\ref{sec15}, then the assumptions that have been made fully
determine the post-data density $p(\mu\,|\,\sigma,x)$ in accordance with equation~(\ref{equ15}) and
with expressions analogous to the ones given in equations~(\ref{equ11}) and~(\ref{equ12}).
Furthermore, the fiducial density of $\sigma$ conditional on $\mu$ that is obtained by using the
strong fiducial argument, i.e.\ the density $f_{S}(\sigma\, |\,\mu, x)$, is defined by the
following expression:
\[
\sigma^2\, |\, \mu, x \sim \mbox{Scale-inv-$\chi^2$} (n,\bm\hat{\sigma}^2)
\]
where $\bm\hat{\sigma}^2 = \sum_{i=1}^{n} (x_i - \mu)^2 / n\hspace{0.1em}$, i.e.\ it is a scaled
inverse $\chi^2$ density function with $n$ degrees of freedom and scaling parameter equal to the
variance estimator $\bm\hat{\sigma}^2$. If we assume, as we will do so from now on, that the sample
variance $s^2$ is equal to 9, then under the other assumptions that have already been made, this
fiducial density $f_{S}(\sigma\, |\,\mu, x)$ can be expressed as:
\[
\sigma^2\, |\, \mu, x \sim \mbox{Scale-inv-$\chi^2$} (\,9,\, 8 + (\mu-2.7)^2\,)
\vspace*{0.5ex}
\]
This density function will therefore be regarded as being the post-data density of $\sigma$
conditional on $\mu$ in the example under discussion.
Again to clarify, it has been assumed in deriving the fiducial density in question that the
fiducial statistic is any sufficient statistic for $\sigma$, e.g.\ the variance estimator
$\bm\hat{\sigma}^2$ just defined.

To illustrate this example, Figure~4 shows some results from running a Gibbs sampler on the basis
of the full conditional post-data densities of $\mu$ and $\sigma$ \pagebreak that have just been
defined, i.e.\ the densities $p(\mu\,|\,\sigma,x)$ and $f_{S}(\sigma\, |\,\mu, x)$, with a uniform
random scanning order of the parameters $\mu$ and $\sigma$, as such a scanning order was defined in
Section~\ref{sec22}.
In particular, the histograms in Figures~4(a) and~4(b) represent the distributions of the values of
$\mu$ and $\sigma$, respectively, over a single run of six million samples of these parameters
generated by the Gibbs sampler after a preceding run of one thousand samples, which were classified
as belonging to its burn-in phase, had been discarded.
The sampling of the density $p(\mu\,|\,\sigma,x)$ was based on the Metropolis algorithm (Metropolis
et al.~1953), while each value drawn from the density $f_{S}(\sigma\, |\,\mu, x)$ was independent
from the preceding iterations.

\begin{figure}[t]
\begin{center}
\noindent
\makebox[\textwidth]{\includegraphics[width=7.25in]{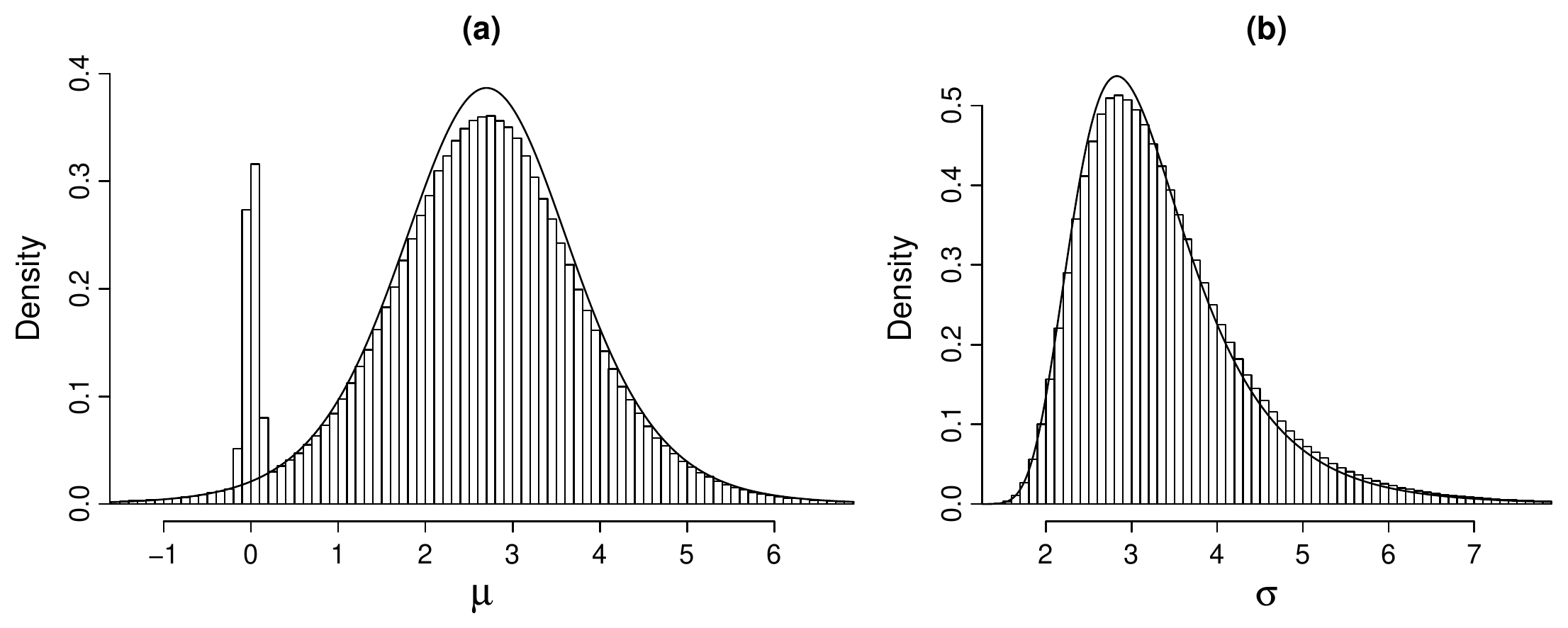}}
\caption{\small{Marginal post-data densities of the mean $\mu$ and standard deviation $\sigma$
of a normal distribution}}
\end{center}
\end{figure}

In accordance with conventional recommendations for evaluating the convergence of Monte Carlo
Markov chains outlined, for example, in Gelman and Rubin~(1992) and Brooks and Roberts~(1998), an
additional analysis was carried out in which the Gibbs sampler was run various times from different
starting points and the output of these runs was carefully assessed for convergence using suitable
diagnostics. This analysis provided no evidence to indicate that the sampler does not have a
limiting distribution, and showed, at the same time, that it would appear to generally converge
quickly to this distribution.

Furthermore, the Gibbs sampling algorithm was run separately with each of the two possible fixed
scanning orders of the parameters, i.e.\ the one in which $\mu$ is updated first and then $\sigma$
is updated, and the one that has the reverse order, in accordance with how a single transition of
such an algorithm was defined in Section~\ref{sec22}, i.e.\ single transitions of the algorithm
incorporated updates of both parameters.
In doing this, no statistically significant difference was found between the samples of parameter
values aggregated over the runs of the sampler in using each of these two scanning orders after
excluding the burn-in phase of the sampler, e.g.\ between the two sample correlations of $\mu$ and
$\sigma$, even when the runs concerned were long.
Taking into account what was discussed in Section~\ref{sec22}, this implies that the full
conditional densities of the limiting distribution of the original Gibbs sampler, i.e.\ the one
with a uniform random scanning order, should be, at the very least, close approximations to the
full conditional densities on which the sampler is based, i.e.\ the post-data densities
$p(\mu\,|\,\sigma,x)$ and $f_{S}(\sigma\, |\,\mu, x)$ defined earlier.

With regard to analysing the same data set $x$, the curves overlaid on the histograms in
Figures~4(a) and~4(b) are plots of the marginal fiducial densities of the parameters $\mu$ and
$\sigma$, respectively, in the case where the joint fiducial density of these parameters is
uniquely defined by the compatible full conditional fiducial densities $f_{S}(\mu\,|\,\sigma, x)$
and $f_{S}(\sigma\, |\,\mu, x)$, which have already been specified in the present section, i.e.\ in
the case where both the parameters $\mu$ and $\sigma$ belong to the set $\theta^{B}$ referred to in
Section~\ref{sec13}. See Bowater~(2018a) for further information about the general type of joint
fiducial density of $\mu$ and $\sigma$ in question, and to add a little more detail, let us clarify
that the marginal fiducial density of $\sigma$ being referred to is defined by:
\vspace*{-0.5ex}
\[
\sigma^2\, |\, x \sim \mbox{Scale-inv-$\chi^2$} (n-1, s^2)
\]
The accumulation of probability mass around the value of zero in the marginal post-data density of
$\mu$ that is represented by the histogram in Figure~4(a), and the fact that the upper tail of the
marginal post-data density of $\sigma$ that is represented by the histogram in Figure~4(b) tapers
down to zero slightly more slowly than the aforementioned marginal fiducial density for $\sigma$
are both clearly artefacts of the strong pre-data opinion that was held about $\mu$.

\vspace*{3ex}
\subsection{Second example: Inference about a relative risk}
\label{sec18}

To go, in a certain sense, completely around the circle of examples considered in the present
paper, let us now try to directly address the problem of inference posed in Example~2 of the
Introduction, i.e.\ that of making inferences about a relative risk $\pi_t/\pi_c$ of a given
adverse event of interest.

For the same reason as given in the description of this example, let us assume that the scenario of
Definition~2 would apply if $\pi_t$ was unknown and $\pi_c$ was known, and also if $\pi_c$ was
unknown and $\pi_t$ was known. In particular, if we define\hspace*{0.1em}
$\mbox{odds}(p)=p/(1-p)$\hspace*{0.05em} then, in the case where $\pi_t$ is the unknown parameter,
the special hypothesis in this scenario will be assumed to be that $\mbox{odds}(\pi_t)$, i.e.\ the
odds of $\pi_t$, lies in the following narrow interval:
\vspace*{-0.5ex}
\[
\mathtt{I}(\pi_c)=(\hspace*{0.1em}\mbox{odds}(\pi_c)/(1+\varepsilon),\hspace*{0.1em}
\mbox{odds}(\pi_c)(1+\varepsilon)\hspace*{0.1em})
\vspace*{-0.5ex}
\]
where $\varepsilon$ is a small positive constant, which is an interval that clearly contains
$\mbox{odds}(\pi_c)$, i.e.\ the odds of $\pi_c$. In a symmetrical way, we will assume that, in the
case where $\pi_c$ is the unknown parameter, the special hypothesis in the scenario of Definition~2
is that $\mbox{odds}(\pi_c)$ lies in the narrow interval\hspace*{0.1em}
$\mathtt{I}(\pi_t)=(\hspace*{0.05em}\mbox{odds}(\pi_t)/(1+\varepsilon),\hspace*{0.1em}
\mbox{odds}(\pi_t)(1+\varepsilon)\hspace*{0.05em})$.
Of course, having a high degree of pre-data belief that $\mbox{odds}(\pi_t) \in \mathtt{I}(\pi_c)$
logically implies that one should have a high degree of pre-data belief that $\mbox{odds}(\pi_c)
\in \mathtt{I}(\pi_t)$, and also we can see that this is consistent with having a high degree of
pre-data belief that \pagebreak the relative risk $\pi_t /\pi_c$ is close to one.
As a result of what has just been assumed, it is clear that, in terms of the notation of
Section~\ref{sec13}, the set of parameters $\theta^{A}$ will contain both the parameters $\pi_t$
and $\pi_c$, while the set $\theta^{B}$ will be empty.

In addition, we will assume that the full conditional fiducial densities $f_S(\pi_t\,|\,\pi_c,x)$
and $f_S(\pi_c\,|\,\pi_t,x)$ that were defined for the general case immediately after
equation~(\ref{equ10}) are both specified such that they do not depend on the conditioning
parameter concerned, and therefore are equivalent to the marginal fiducial densities
$f_S(\pi_t\,|\,x)$ and $f_S(\pi_c\,|\,x)$, respectively, which is a natural assumption to make
given the independence of the data between the treatment and control groups. More specifically, let
us suppose that each of these fiducial densities is defined in the same way as the fiducial density
$f_{S}(\pi\,|\,x)$ was defined in Section~\ref{sec16}, i.e.\ on the basis of the LPD
function\hspace*{0.05em} $\omega_L (\pi)$ given in equation~(\ref{equ16}).

Although it would not of course be appropriate, given what was known about $\pi_t$ and $\pi_c$
before the data were observed, to use the joint fiducial density of $\pi_t$ and $\pi_c$ that is
defined by the marginal fiducial densities $f_S(\pi_t\,|\,x)$ and $f_S(\pi_c\,|\,x)$ as a post-data
density function of $\pi_t$ and $\pi_c$ in the present example, it would nevertheless be
interesting to see what the marginal density of the relative risk $\pi_t/\pi_c$ over this joint
fiducial density would look like.
Therefore, the histogram in Figure~5 represents this marginal density of $\pi_t/\pi_c$ for the case
where, in terms of the notation established for the present ex\-am\-ple in the Introduction, the
observed counts are specified as follows: $e_t=6$, $n_t=20$, $e_c=18$ and $n_c=30$. This histogram
was constructed on the basis of a sample of three million independent random values drawn from the
marginal fiducial density of $\pi_t/\pi_c$ \linebreak concerned.

\begin{figure}[!t]
\begin{center}
\includegraphics[width=4in]{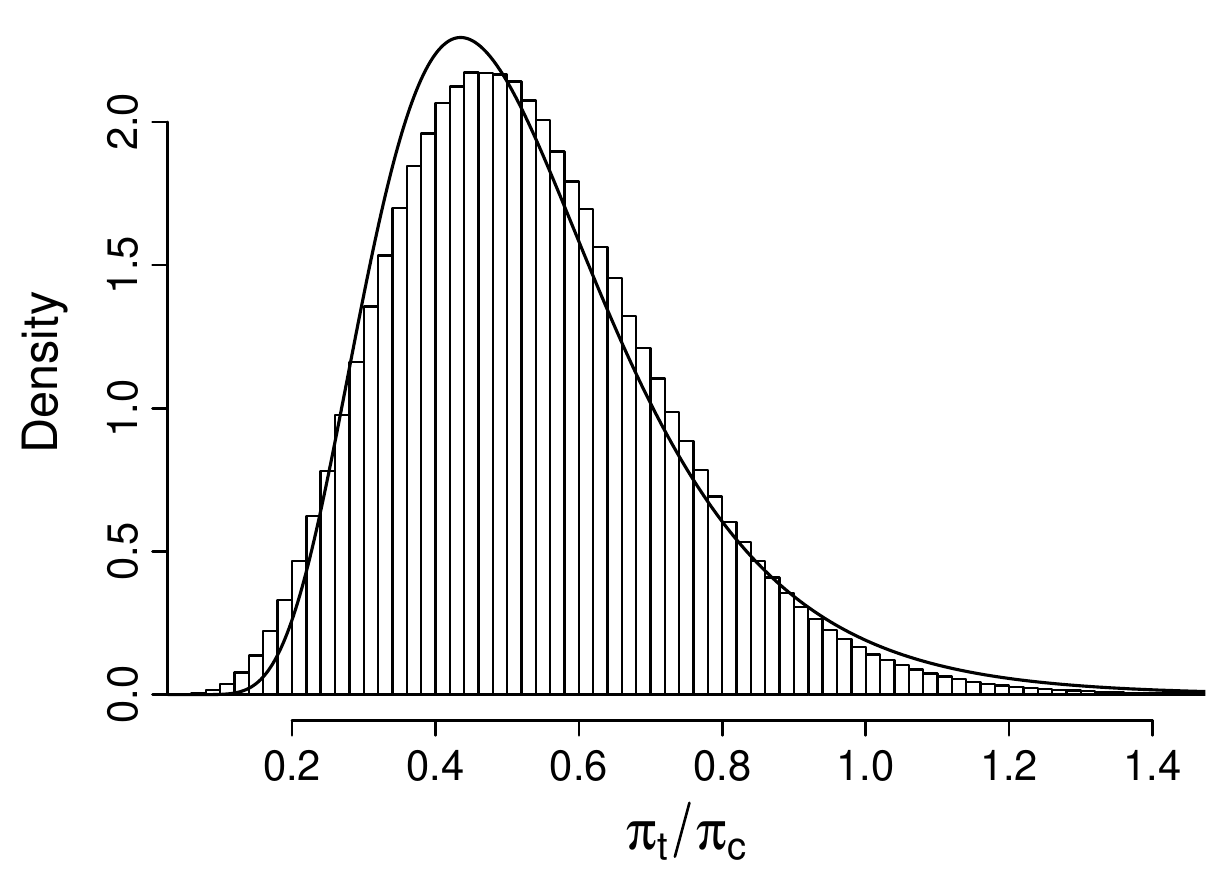}
\caption{{\small Histogram representing a marginal fiducial density of the relative risk
$\pi_t/\pi_c$}}
\end{center}
\end{figure}

By contrast, it is standard practice to form (Neyman-Pearson) confidence intervals for any given
relative risk by approximating the density function of the logarithm of the sample relative risk by
a normal density function centred at the logarithm of the true relative risk, with the usual
estimate of the variance of this density function treated as though it is the true value of this
variance (see for example Altman~1991).
For this reason, a curve has been overlaid on the histogram in Figure~5 representing a confidence
density function of the relative risk $\pi_t/\pi_c$ that, with regard to the data of current
interest, was constructed in the usual way on the basis of the type of confidence interval being
referred to by allowing the coverage probability of this relative risk to vary (see for example
Efron~1993 for further clarification about how this type of function is defined).
It can be clearly seen that this confidence density function is very different from the density
function of $\pi_t/\pi_c$ that is depicted by the histogram in this figure, which indicates the
inadequacy of the approximate confidence intervals in question, even if nothing or very little was
known about the parameters $\pi_t$ and $\pi_c$ before the data were observed.

To simplify proceedings, let us from now on assume both that $\varepsilon=0.08$ and that the event
counts $e_t$, $n_t$, $e_c$ and $n_c$ are equal to the values that have just been given. Under these
assumptions, if for a given value of $\pi_c$\hspace*{0.05em}, the following condition holds
\vspace*{1ex}
\begin{equation}
\label{equ28}
\beta_0 = \sum_{y=0}^6 \binom{20}{y} (\pi_{t0})^{y}(1-\pi_{t0})^{20-y}\hspace*{0.2em}
\leq\hspace*{0.2em} \sum_{y=6}^{20} \binom{20}{y} (\pi_{t1})^{y}(1-\pi_{t1})^{20-y} = \beta_1
\end{equation}
where $\pi_{t0} = \mbox{odds}^{-1}(\mbox{odds}(\pi_c)/(1.08))$ and
$\pi_{t1} = \mbox{odds}^{-1}(\mbox{odds}(\pi_c)(1.08))$,
then in determining the post-data density $p(\pi_t\,|\,\pi_c,x)$, the hypotheses $H_{P}$ and
$H_{S}$ would be defined as:
\vspace*{-2.5ex}
\begin{gather}
H_{P}: \pi_t \geq \pi_{t0} \label{equ29}\\[-0.5ex]
H_{S}: \rho (E_t^* \leq 6) \leq \beta_0 \label{equ30}
\end{gather}
\par \vspace*{-0.5ex} \noindent
while otherwise, these hypotheses would have the definitions:
\vspace*{-0.5ex}
\begin{gather}
H_{P}: \pi_t \leq \pi_{t1} \label{equ31}\\[-0.5ex]
H_{S}: \rho (E_t^* \geq 6) \leq \beta_1 \label{equ32}
\end{gather}
where $\beta_0$ and $\beta_1$ are the one-sided P-values defined in equation~(\ref{equ28}) and
where in both definitions of the hypothesis $H_{S}$, the random variable $E_t^*$ is assumed to be
the number of patients in an additional as-yet-unrecruited group of
$n_t\hspace*{-0.1em}=\hspace*{-0.1em}20$ patients who would experience the given adverse event of
interest when receiving the same drug that was administered to the patients in the treatment group,
i.e.\ drug B.

Clearly, since the one-sided P values $\beta_0$ and $\beta_1$ in the two versions of the hypothesis
$H_{S}$ in equations~(\ref{equ30}) and~(\ref{equ32}) depend on the value of $\pi_c$, a PDO curve is
going to be useful in assigning values to the probability that any given version of this hypothesis
is true.
Moreover, it can be seen from equation~(\ref{equ28}) that the way the hypotheses $H_{P}$ and
$H_{S}$ are going to be defined in any given case out of the two ways of defining these hypotheses
given in equations~(\ref{equ29}), (\ref{equ30}), (\ref{equ31}) and~(\ref{equ32}) will also depend
on the value of $\pi_c$. For this reason, to completely describe, in all circumstances of interest,
what probability should be given to the hypothesis $H_{S}$, i.e.\ the probability $\alpha$, we
would require the information that is conveyed by two separate PDO curves, each one corresponding
to one of the two definitions of this hypothesis.
However, for the purpose of giving an example, it will be assumed that these two PDO curves are in
fact the same, which actually is a reasonably justifiable assumption to make.
In particular, we will assume that the single PDO curve being referred to has the simple algebraic
form:
\vspace*{-0.5ex}
\begin{equation}
\label{equ19}
\alpha = ((0.92)\beta)^{\hspace*{0.1em}0.6}
\vspace*{-0.5ex}
\end{equation}
where $\beta$ is the P value of interest, i.e.\ the value $\beta_0$ or $\beta_1$.

This PDO curve is represented by the solid curve in Figure~6(a).
The two lowest dashed curves in this figure represent, on the other hand, what the PDO curve under
discussion would need to be so that the probability $\alpha$ equals the probability that would be
assigned to the hypothesis $H_{S}$ by the fiducial density $f_{S}(\pi_t\,|\,x)$ defined earlier.
Each of these curves corresponds to one of the two definitions of the hypothesis $H_{S}$. Of
course, we would expect any choice for the PDO curve in question to be always higher than these two
dashed curves, which is the case, as can be seen, for the PDO curve that has been proposed.

\begin{figure}[!t]
\begin{center}
\makebox[\textwidth]{\includegraphics[width=7.25in]{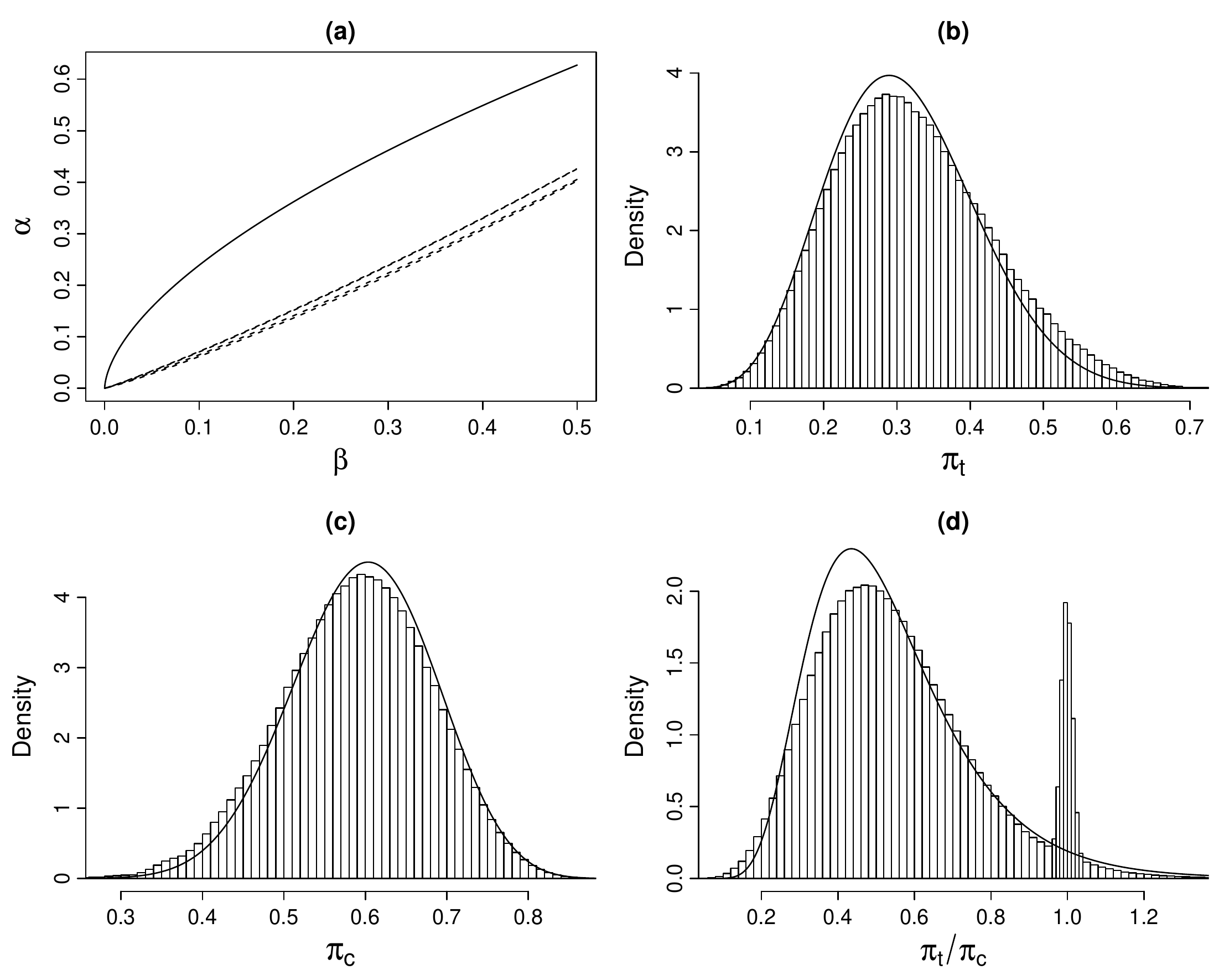}}
\caption{\small{Graph (a) shows the PDO curve used when either $\pi_t$ or $\pi_c$ is treated as the
only unknown parameter. The histograms in graphs (b) to (d) represent marginal post-data densities
of the binomial proportions $\pi_t$ and $\pi_c$, and the relative risk $\pi_t/\pi_c$,
respectively.}}
\end{center}
\end{figure}

In determining the post-data density $p(\pi_c\,|\,\pi_t,x)$, the hypotheses $H_{P}$ and $H_{S}$
would be defined in a similar way to how they have just been specified. Even though it is now
$\pi_c$ rather than $\pi_t$ that is the unknown parameter, it will be assumed that the PDO curves
that are associated with the two definitions of the hypothesis $H_{S}$ that apply in this case are
again equal to the single PDO curve given in equation~(\ref{equ19}), which, taking into account
especially the discrete nature of the data, is a slightly unsophisticated, but nonetheless,
adequate assumption to make for the purposes of giving an example.

The two highest short-dashed curves in Figure~6(a), which in fact appear to be almost a single
long-dashed curve because they are so close together, represent what the PDO curve in this case,
i.e.\ in the case where $\pi_t$ is known and $\pi_c$ is unknown, would need to be so that the
probability $\alpha$ equals the probability that would be assigned to the hypothesis $H_{S}$ by the
fiducial density $f_{S}(\pi_c\,|\,x)$ defined earlier.
As before, each curve corresponds to one of the definitions of the hypothesis $H_{S}$. These two
curves are clearly quite close to the other two dashed curves in this figure, but reassuringly a
long way below the proposed PDO curve.

Given the assumptions that have already been made, we only need to specify the density functions
$h(\pi_t)$ and $h(\pi_c)$ that are required by equation~(\ref{equ14}) in order to be able to fully
determine the full conditional post-data densities $p(\pi_t\,|\,\pi_c,x)$ and
$p(\pi_c\,|\,\pi_t,x)$ according to equation~(\ref{equ15}) and to expressions identical or
analogous to the ones given in equations~(\ref{equ11}) and~(\ref{equ12}).
With this aim in mind, let us therefore define the density functions $h(\pi_t)$ and $h(\pi_c)$ such
that $\log(\mbox{odds}(\pi_t))$ and $\log(\mbox{odds}(\pi_c))$ have beta density functions on the
intervals obtained by converting, respectively, the intervals $\mathtt{I}(\pi_c)$ and
$\mathtt{I}(\pi_t)$ defined earlier to the logarithmic scale, with both shape parameters of these
density functions equal to 4.

To illustrate this example, Figure~6 shows some results from running a Gibbs sampler on the basis
of good approximations to the full conditional post-data densities of $\pi_c$ and $\pi_t$ that have
just been defined, i.e.\ the densities $p(\pi_t\,|\,\pi_c,x)$ and $p(\pi_c\,|\,\pi_t,x)$, with a
uniform random scanning order of the parameters $\pi_t$ and $\pi_c$.
In particular, the histograms in Figures~6(b) to~6(d) represent the distributions of the values of
$\pi_t$, $\pi_c$ and the relative risk $\pi_t/\pi_c$ respectively, over a single run of six million
samples of the parameters $\pi_t$ and $\pi_c$ generated by the Gibbs sampler after a preceding run
of two thousand samples were discarded due to these samples being classified as belonging to its
burn-in phase.
The sampling of both the densities $p(\pi_t\,|\,\pi_c,x)$ and $p(\pi_c\,|\,\pi_t,x)$ was based on
the Metropolis algorithm.

Similar to what was done in relation to the example considered in the last section, an additional
analysis was carried out in which the Gibbs sampler was run various times from different starting
points and the output of these runs was carefully assessed for convergence using appropriate
diagnostics. Again, this type of analysis provided no evi\-dence to suggest that the sampler does
not have a limiting distribution.

Furthermore, after excluding the burn-in phase of the sampler, no statistically significant
difference was found between the samples of parameter values aggregated over the runs of the
sampler in using each of the two fixed scanning orders of the parameters $\pi_t$ and $\pi_c$ that
are possible, with a single transition of the sampler defined in the same way \linebreak as in the
example outlined in the previous section, even when the runs concerned were long.
Therefore, taking into account what was discussed in Section~\ref{sec22}, the full conditional
densities of the limiting distribution of the original random-scan Gibbs sampler should be, at the
very least, close approximations to the full conditional densities on which the sampler is based,
i.e.\ the post-data densities $p(\pi_t\,|\,\pi_c,x)$ and $p(\pi_c\,|\,\pi_t,x)$ defined earlier.

As already mentioned, there was an approximate aspect to how the Gibbs sampling algorithm generated
random values from the densities $p(\pi_t\,|\,\pi_c,x)$ and $p(\pi_c\,|\,\pi_t,x)$ in \linebreak
question.
However this was in fact simply due to the fiducial densities $f_{S}(\pi_t\,|\,x)$ and
$f_{S}(\pi_c\,|\,x)$, which enter into the definitions of the post-data densities
$p(\pi_t\,|\,\pi_c,x)$ and $p(\pi_c\,|\,\pi_t,x)$, being approximated, respectively, by the
posterior densities of $\pi_t$ and $\pi_c$ that are based on the use of the corresponding Jeffreys
prior for these parameters, which can be recalled is the same type of approximation as was used in
Section~\ref{sec16}.
Similar to earlier, additional simulations showed that, for the data set of current interest, using
this method to approximate the two fiducial densities concerned was very adequate.

Under the same type of approximation, the curves that are plotted in Figures~6(b) and~6(c)
represent the fiducial densities $f_{S}(\pi_t\,|\,x)$ and $f_{S}(\pi_c\,|\,x)$ respectively.
It can be seen that relative to these fiducial densities, the marginal post-data density of $\pi_t$
that is represented by the histogram in Figure~6(b) could be described as being drawn towards the
proportion $e_c/n_c=0.6$, i.e.\ the sample proportion of patients in the control group who
experienced the adverse event of interest, which is especially apparent in the upper tail of this
density, while the marginal post-data density of $\pi_c$ that is represented by the histogram in
Figure~6(c) could be described as being drawn towards this sample proportion in the treatment
group, i.e.\ the value $e_t/n_t=0.3$, which is especially \linebreak apparent in the lower tail of
this density. These characteristics of the marginal post-data densities depicted by the histograms
in question would of course be expected given the nature of the pre-data opinion about $\pi_t$ and
$\pi_c$ that was incorporated into the construction of the joint post-data density of $\pi_t$ and
$\pi_c$ to which these marginal densities correspond.

The curve that is plotted in Figure 6(d), by contrast, represents the confidence density function
of the relative risk $\pi_t/\pi_c$ that was referred to earlier, and which is also depicted by the
curve in Figure~5.
As can be seen, it is very different from the marginal post-data density of this relative risk that
is represented by the histogram in this figure. Of course, the accumulation of probability mass
around the value of one for $\pi_t/\pi_c$ in this latter density function is clearly an artefact of
the strong pre-data opinion that was held about the parameters concerned.

\vspace*{3ex}
\section{A closing remark}

Taking into account all that was discussed in the Introduction, observe that, even to attempt to
construct adequate Bayesian solutions to problems of inference that are loosely similar to the type
of problem that has been of interest in the present paper would require the elicitation of $k$ full
conditional prior density functions, where $k$ of course is the total number of parameters in the
sampling model.
On the other hand, applying the bispatial-fiducial method put forward in the preceding sections to
these problems would require, either automatically or under what was assumed in
Section~\ref{sec18}, the specification of $m$ post-data opinion (PDO) curves, assuming that the set
$\theta^{A}$, as defined in \linebreak Section~\ref{sec13}, contains $m$ parameters.
Therefore, since it must be the case that $m \leq k$, it can be argued that this latter method does
not carry an extra burden in terms of the required incorporation of pre-data knowledge about model
parameters into the inferential process relative to the Bayesian method.
Moreover, the case can reasonably be made that, given that they are formed on the basis of the data
set that has actually been observed, these PDO curves will generally be easier to determine than
the prior density functions in question, and above all, of course, by using the bispatial-fiducial
method, we can obtain a direct solution to the precise problem that we actually have been concerned
with rather than to a related, but quite distinct, version of this problem that may, perhaps, be
more naturally addressed using the Bayesian approach to inference.

\vspace*{5ex}
\noindent
{\bf References}

\begin{description}

\setlength{\itemsep}{1ex}

\item[] Altman, D. G. (1991).\ \emph{Practical statistics for medical research}, Chapman and Hall,
London.

\item[] Bowater, R. J. (2017).\ A defence of subjective fiducial inference.\ \emph{AStA Advances in
Statistical Analysis}, {\bf 101}, 177--197.

\item[] Bowater, R. J. (2018a).\ Multivariate subjective fiducial inference.\ \emph{arXiv.org
(Cornell University), Statistics}, arXiv:1804.09804.

\item[] Bowater, R. J. (2018b).\ On a generalised form of subjective probability.\ \emph{arXiv.org
(Cornell University), Statistics}, arXiv:1810.10972.

\item[] Bowater, R. J. (2019a).\ Organic fiducial inference.\ \emph{arXiv.org (Cornell
University), Sta\-tis\-tics}, arXiv:1901.08589.

\item[] Bowater, R. J. and Guzm\'an-Pantoja, L. E. (2019b).\ Bayesian, classical and hybrid methods
of inference when one parameter value is special.\ \emph{Journal of Applied Statistics}, {\bf 46},
1417--1437.

\item[] Brooks, S. P. and Roberts, G. O. (1998).\ Convergence assessment techniques for Markov
chain Monte Carlo.\ \emph{Statistics and Computing}, {\bf 8}, 319--335.

\item[] Efron, B. (1993).\ Bayes and likelihood calculations from confidence intervals.\
\emph{Bio\-met\-rika}, {\bf 80}, 3--26.

\item[] Fisher, R. A. (1935).\ The fiducial argument in statistical inference.\ \emph{Annals of
Eugenics}, {\bf 6}, 391--398.

\item[] Fisher, R. A. (1956).\ \emph{Statistical Methods and Scientific Inference}, 1st ed., Hafner
Press, New York [2nd ed., 1959; 3rd ed., 1973].

\item[] Gelfand, A. E. and Smith, A. F. M. (1990).\ Sampling-based approaches to calculating
marginal densities.\ \emph{Journal of the American Statistical Association}, {\bf 85}, 398--409.

\item[] Gelman, A. and Rubin, D. B. (1992).\ Inference from iterative simulation using multiple
sequences.\ \emph{Statistical Science}, {\bf 7}, 457--472.

\item[] Geman, S. and Geman, D. (1984).\ Stochastic relaxation, Gibbs distributions and the
Bayesian restoration of images.\ \emph{IEEE Transactions on Pattern Analysis and Machine
Intelligence}, {\bf 6}, 721--741.

\item[] Metropolis, N., Rosenbluth, A. W., Rosenbluth, M. N., Teller, A. H. and Teller, E. (1953).\
Equation of state calculations by fast computing machines.\ \emph{Journal of Chemical Physics},
{\bf 21}, 1087--1092.

\end{description}

\end{document}